\documentclass[amsmath, amsthm, amssymb, floatfix, reprint, twocolumn, notitlepage, nofootinbib, 10pt]{revtex4-1}

\newtheorem{theorem}{Theorem}
\newtheorem{lemma}{Lemma}

\newtheorem{assumption}{Assumption}
\newtheorem{definition}{Definition}

\usepackage[utf8]{inputenc}
\usepackage{microtype,bm,bbm,graphicx,booktabs,times}
\usepackage[usenames,dvipsnames]{xcolor}
\usepackage{setspace}
\newcommand\blfootnote[1]{%
  \begingroup
  \renewcommand\thefootnote{}\footnote{#1}%
  \addtocounter{footnote}{-1}%
  \endgroup
}
\selectfont{}

\usepackage{hyperref}
\usepackage{braket}
\usepackage{subfiles} 
\usepackage{upgreek}
\usepackage{color}
\hypersetup{colorlinks,allcolors=black,linktoc=all}
\usepackage[capitalize]{cleveref}
\crefformat{section}{Sec.~#2#1#3}

\newcommand{\edit}[1]{{\color{black} #1}}

\newcommand{\norm}[1]{\left\Vert{#1}\right\Vert}

\begin{document}
\title{Convergence guarantees for discrete mode approximations to non-Markovian quantum baths}
\author{Rahul Trivedi$^{1, 2}$}
\email{rahul.trivedi@mpq.mpg.de}
\author{Daniel Malz$^{1, 2}$}
\author{J.~Ignacio Cirac$^{1, 2}$}
\address{$^1$Max-Planck-Institut für Quantenoptik, Hans-Kopfermann-Str.~1, 85748 Garching, Germany.\\
$^2$Munich Center for Quantum Science and Technology (MCQST), Schellingstr. 4, D-80799 Munich, Germany.}

\date{\today}

\begin{abstract}
Non-Markovian effects are important in modeling the behavior of open quantum systems arising in solid-state physics, quantum optics as well as in study of biological and chemical systems. The non-Markovian environment is often approximated by discrete bosonic modes, thus mapping it to a Lindbladian or Hamiltonian simulation problem. While systematic constructions of such modes have been previously proposed, the resulting approximation lacks rigorous and general convergence guarantees. In this letter, we show that under some physically motivated assumptions on the system-environment interaction, the finite-time dynamics of the non-Markovian open quantum system computed with a sufficiently large number of modes is guaranteed to converge to the true result. Furthermore, we show that this approximation error typically falls off polynomially with the number of modes. Our results lend rigor to classical and quantum algorithms for approximating non-Markovian dynamics.
\end{abstract}

\maketitle
\newpage

Quantum systems invariably interact with their environment, and any simulation technique used to model their behaviour needs to capture this interaction. Traditionally, such interactions are analyzed within the Markovian approximation, wherein the system dynamics is described by the Lindbladian master equation \cite{breuer2002theory}. However, a number of quantum systems arising in solid-state physics \cite{finsterholzl2020nonequilibrium, chin2011generalized, groeblacher2015observation, de2008matter}, quantum optics \cite{calajo2019exciting, andersson2019non, gonzalez2019engineering, aref2016quantum, leonforte2021vacancy} as well as quantum biology and chemistry \cite{ishizaki2005quantum, chin2012coherence, ivanov2015extension, caycedo2021exact} cannot be modeled accurately within the Markovian approximation and the non-Markovian nature of the environment needs to be explicitly taken into account.

Simulating non-Markovian open quantum systems is difficult since it is usually not possible to formulate a dynamical equation for the system state from a given physical model of the system-environment interaction. While it is generically expected that non-Markovian open quantum systems satisfy a master equation of the Nakajima-Zwanzig form \cite{xu2018convergence, ivanov2015extension}, or the time-convolutionless form \cite{smirne2010nakajima, pereverzev2006time,kidon2015exact}, it is usually hard to obtain an explicit form of such a master equation except when the system is only weakly coupled to its environment \cite{vacchini2010exact, schroder2007reduced, timm2011time, mukamel1978statistical}. Even though significant progress has been made in utilizing influence functionals or their variants for describing and simulating non-Markovian dynamics \cite{albeverio2007rigorous, makarov1994path, smith1987generalized, whalen2017open, grimsmo2015time, jorgensen2019exploiting,del2018tensor,strathearn2018efficient, rosenbach2016efficient}, the worst-case classical complexity of these approaches increases exponentially with time. An alternative approach is to identify and track a set of discrete modes that approximate the environment \cite{huh2014linear, chin2010exact, woods2014mappings}. This maps the simulation of the non-Markovian open quantum system to a larger Hamiltonian or Lindbladian simulation problem, which can be solved using standard classical \cite{vidal2004efficient,daley2004time,banuls2009matrix,daley2009atomic,verstraete2004matrix} or quantum algorithms \cite{berry2014exponential, low2019hamiltonian, berry2015hamiltonian, zanardi2016dissipative,chenu2017quantum,cleve2016efficient}.

For gaussian bosonic environments \cite{budini2000non,megier2018parametrization}, there are two prominent approaches to identifying these discrete modes. First is to use the Lorentzian pseudomode theory \cite{tamascelli2018nonperturbative, pleasance2020generalized, mazzola2009pseudomodes, dalton2001theory, garraway2006theory, dalton2012quasimodes}, wherein the spectral density function of the non-Markovian environment is approximated by a finite sum of Lorentzians, each of which corresponds to an individual bosonic mode coupled to Markovian reservoir. The second method is to use star-to-chain transformation \cite{chin2010exact, woods2014mappings}, which uses the Lanczos iteration to identify a 1D chain of discrete bosonic modes with nearest neighbour couplings that approximate the environment and map the problem of computing non-Markovian quantum dynamics to a Hamiltonian simulation problem. While a systematic construction of these approximations have been given, their convergence properties are less well understood. For instance, classes of models where the Lorentzian pseudomode description is exact are known \cite{tamascelli2018nonperturbative, pleasance2020generalized}, but it is unknown if it can efficiently approximate a non-Markovian environment. More attention has been paid towards investigating the convergence properties of the star-to-chain transformation --- convergence guarantees have been provided for models with a finite Lieb-Robinson velocity \cite{woods2015simulating, gualdi2013renormalization}, or for bounded memory kernels \cite{mascherpa2017open} that can possibly have an infinite Lieb-Robinson velocity \cite{eisert2009supersonic}. However, these analyses do not extend to models with distributional (and hence unbounded) memory kernels, such as those commonly encountered in non-Markovian quantum optical systems \cite{calajo2019exciting, fan2010input}.

In this letter, we provide general and rigorous convergence guarantees for discrete-mode approximations of non-Markovian gaussian bosonic environments. We show that for a wide and physically-motivated class of non-Markovian models, both the Lorentzian pseudomode approximation and the star-to-chain transformation is guaranteed to converge and the approximation error falls off polynomially with the number of pseudomodes. Our results not only provide the first set of rigorous convergence guarantees for the pseudomode approximation, but also extend the convergence guarantees for the star-to-chain transformation to non-Markovian systems described by a distributional memory kernel, such as those commonly encountered in quantum optics \cite{calajo2019exciting, fan2010input}.

In order to perform this convergence study, there are several theoretical challenges that our work resolves. One of the key issues with analyzing models of non-Markovian environments is that the environment can support arbitrarily large energies and the quantum system often has non-vanishing couplings with high energy environment modes  \cite{calajo2019exciting, fan2010input}. We identify a set of physically motivated sufficient mathematical conditions on the system-environment dynamics that allow for rigorously neglecting the high energy modes in the environment while studying finite-time dynamics. While this straightforwardly follows for problems where the system-environment couplings vanish at high energies, our analysis also includes distributional memory kernels. Combining this with an analysis of the Lorentzian pseudo-mode approximation and the star-to-chain transformation within a truncated environment-energy window, we provide convergence guarantees as well estimates of the rate of convergence for both of these methods.

We consider an open quantum system model, with a $d$-dimensional quantum system with hilbert space $\mathcal{H}_S = \mathbb{C}^d$ (referred to as a `local system') interacting with a gaussian environment whose Hilbert space, $\mathcal{H}_E$ is assumed to be a fock space over $\text{L}^2(\mathbb{R})$. We denote by $a_\omega$ the annihilation operator for this Fock space and consider Hamiltonians of the form
\begin{align}\label{eq:hamiltonian}
H(t) =& H_S(t) + \int_{-\infty}^{\infty} \omega a^\dagger_\omega a_\omega d\omega + \nonumber\\
& \int_{-\infty}^\infty \bigg( v(\omega) a_\omega L^\dagger + v^*(\omega) a^\dagger_\omega L\bigg) d\omega,
\end{align}
where $L: \mathcal{H}_S \to \mathcal{H}_S$ is the operator describing the coupling of the system with the environment, and $v$ is the frequency dependent coupling function between the environment and the system. We point out that the Hamiltonian in Eq.~\ref{eq:hamiltonian} is only a provably (essentially) self-adjoint operator if $v \in \text{L}^2(\mathbb{R})^\dagger$\blfootnote{$^\dagger$A map $v:\mathbb{R}\to \mathbb{C}$ is in $\textrm{L}^2(\mathbb{R})$ if $v$ is square integrable with respect to the Lesbesgue measure}. However, several problems in quantum optics (e.g. systems with point coupling) are described by $v \in \textrm{C}_b^\infty(\mathbb{R}) \cap \mathcal{S}'(\mathbb{R})^{\dagger\dagger}$\blfootnote{$\dagger\dagger$A map $v:\mathbb{R} \to \mathbb{C}$ is in $\textrm{C}_b^\infty(\mathbb{R})$ if $v$ is smooth and bounded. For $v \in \textrm{C}_b^\infty(\mathbb{R})$, $\norm{v}_\infty = \sup_{\omega \in \mathbb{R}} |v(\omega)|$. We will denote the space of tempered distributions over $\mathbb{R}$ by $\mathcal{S}'(\mathbb{R})$ --- $\textrm{C}^\infty_b(\mathbb{R})\cap \mathcal{S}'(\mathbb{R})$ is the space of tempered distributions that correspond to multiplication with bounded and smooth functions.} that are tempered distributions and still result in well defined local system dynamics. Throughout this paper, we will be interested in approximating the dynamics of the reduced state of the local system --- \edit{for simplicity, we restrict ourselves to the case where the environment is initially in a vacuum state, extensions of the main results of this paper to initially excited environment states is provided in the supplement \cite{supp}.}

First we outline two physically motivated assumptions that we make on the model under consideration that are sufficient conditions to allow us to neglect large environment frequencies. \edit{The first assumption makes mathematically precise the expectation that the effective time-domain kernel, $K_v$ corresponding to the coupling function $v \in \text{C}_b^\infty(\mathbb{R}) \cap \mathcal{S}'(\mathbb{R})$, defined by
\begin{align*}
K_v(t) = \int_{-\infty}^\infty |v(\omega)|^2 e^{-i\omega t} d\omega,
\end{align*}
is approximable by its restriction to a finite frequency window. Since $K$ can in general be a distribution, we first introduce a distributional quasi-norm to quantify this approximation error. \\  \ \\

\begin{definition}
Let $v \in \mathbb{C}_b^\infty(\mathbb{R})$, and let $K_v$ be its corresponding kernel, then for $t > 0$, define a quasi-norm $\ell(K_v, t)$ of $K_v$ via\blfootnote{ A map $f: [a, b]^2 \to \mathbb{C}$ will be called symmetric if $\forall s_1, s_2 \in [a, b]: \ f(s_1, s_2) = f^*(s_2, s_1)$.  We will denote by $\text{AC}_\text{sym}([a, b]^2)$ the space of symmetric, continuous functions $f:[a, b]^2 \to \mathbb{C}$ which satisfy the following two conditions
\begin{enumerate}
\item $f$ is absolutely continuous in either of its arguments i.e. $\forall y \in [a, b],$ the map $\varphi_y: [a, b] \to \mathbb{C}$ defined by $\varphi_y(x) = f(x, y)$ is absolutely continuous.
\item For $y \in_{a.e.}[a, b]$, $\partial_y f(x, y)$ exists $\forall x \in [a, b]$ and the map $\varphi_y^d: [a, b] \to \mathbb{C}$ defined by $\varphi_y^d(x) = \partial_y f(x, y)$ is absolutely continuous.
\end{enumerate}
We note that the partial derivatives of $f$ with respect to either and both of its arguments, $\partial_x f$, $\partial_y f$, $\partial^2_{x,y} f$, exists almost everywhere on $[a, b]^2$. Finally, we define a norm $\norm{\cdot}_{[a, b]^2}^{\mathcal{S}}: \text{AC}_\text{sym}([a, b]^2) \to \mathbb{R}^+$ such that $\forall  f \in \text{AC}_\text{sym}([a, b]^2)$
\begin{align*}
\norm{f}^{\mathcal{S}}_{[a, b]^2} &=  \sup_{x, y \in [a, b]} |f(x, y)| + \nonumber \\ &(b - a) \bigg(\underset{x, y \in [a, b]}{\text{esssup}} \big|\partial_x f(x, y)\big| + \underset{{x, y \in [a, b]}}{\text{esssup}} \big|\partial_y f(x, y)\big|\bigg) + \nonumber\\ &(b-a)^2\underset{{x, y \in [a, b]}}{\text{esssup}} \big|\partial^2_{x, y} f(x, y)\big| .
\end{align*}
A symmetric map $K$ is also positive if
\[
\forall g \in \text{C}^0([a, b]): \ \int_{s_1, s_2 = a}^b K(s_1, s_2) g^*(s_1) g(s_2) ds_1 ds_2 \geq 0.
\]
We define $\text{AC}_\text{sym}^{\succeq 0}([a, b]^2) = \{f \in \text{AC}_\text{sym}([a, b]^2) | f \text{ is positive.}\}$
}
\begin{widetext}
\begin{align*}
&\ell(K_v, t) = \sup_{\substack{{f \in \textnormal{AC}_\textnormal{sym}^{\succeq 0}}([0, t]^2) ^{\ddagger}\\ f \neq 0}} \frac{1}{{\norm{f}_{[0, t]^2}^\mathcal{S}}}{\int_{s_1, s_2 = 0}^t K_v(t - s_1) K^*_v(t - s_2) f(s_1, s_2) ds_1 ds_2}.
\end{align*}
\end{widetext}
\end{definition}
}
\edit{We note that our definition of this quasi-norm involves two applications of the kernel on a test function, which itself is a function of two time indices --- this choice of the quasi-norm is natural since we are interested in quantifying the back-action of particles emitted into the environment on the local system dynamics, with there being two time indices needed to describe the reduced state of each particle in the environment.}  \\ \ \\

\edit{\begin{assumption} The coupling function $v \in \mathbb{C}_b^\infty(\mathbb{R})$ is such that there is a function $V(\omega_c, t)$ which vanishes as $\omega_c\to \infty \ \forall t \geq 0$ and 
\begin{align*}
\ell(K_v - K_{v_{\omega_c}}, t) \leq V(\omega_c, t)
\end{align*}
where $v_{\omega_c}(\omega) = v(\omega)$ if $|\omega| \leq \omega_c$ and otherwise 0.
\end{assumption}}
\edit{By ensuring that the kernel is distributionally approximable within a finite frequency window, assumption 1 ensures that the model described by Eq.~\ref{eq:hamiltonian} does not suffer from an ultraviolet divergence arising due to the environment being able to support arbitrarily high frequencies \cite{barut1983nonperturbative}.} A number of models commonly considered in practice do satisfy this assumption
\begin{enumerate}
\item \emph{Environments with square-integrable coupling functions,} i.e.~$v\in \textrm{C}_b^\infty(\mathbb{R}) \cap \textrm{L}^2(\mathbb{R})$ satisfy assumption 1 (proposition 1 in the supplement \cite{supp}). Physically important examples of such environments include environments with a Lorentzian coupling function which is typically used to model an atomic system interacting with an optical cavity \cite{garraway1997nonperturbative}.
\item \emph{Markovian environments}, which correspond to a frequency independent coupling function $v(\omega) = v_0$ \cite{fan2010input}, also satisfy assumption 1 (proposition 2 in the supplement \cite{supp}). 
\item \emph{Environments modelling retardation effects,} described by coupling functions of the form $v(\omega) = \sum_{i=1}^M v_i e^{i\omega \tau_i}$ are commonly used to model retardation effects \cite{cilluffo2020collisional}. These environments also satisfy the conditions of assumption 1 (proposition 2 in the supplement \cite{supp}).
\end{enumerate}

\edit{We point out that while the kernel might be approximable within a finite frequency window, due to the infinite-dimensional nature of the environment's Hilbert space, it does not immediately follow that the reduced state of the local system is also similarly approximable.} To proceed further, we introduce a second assumption that ensures physically meaningful joint system-environment dynamics. To state this assumption mathematically, we introduce the $N-$point Green's function of the localized system.

\begin{definition}[$N-$point Green's function]
$\forall s \in [0, t]^N$, $G_N(s; t) \in \mathfrak{L}(\mathcal{H}_S)^{\S}$\blfootnote{$\S$ For a Hilbert space $\mathcal{H}$, $\mathfrak{L}(\mathcal{H})$ is the set of operators $T:\mathcal{H}\to \mathcal{H}$ which are bounded. For $T\in \mathfrak{L}(\mathcal{H}_S)$, $\norm{\cdot}$ is the usual operator norm defined by $\norm{T} = \sup_{\ket{\psi}\in \mathcal{H}_S \setminus\{0\}} \norm{T\ket{\psi}} / \norm{\ket{\psi}}$.} via
\begin{align}
G_N(s; t) = \bra{\textnormal{vac}} U(t, 0)\mathcal{T}\bigg[\prod_{i=1}^N L(s_i)\bigg] \ket{\textnormal{vac}},
\end{align}
where $U(\tau_1, \tau_2)$ is the propagator corresponding to the Hamiltonian in Eq.~\ref{eq:hamiltonian} and $L(\tau) = U(0, \tau) L U(\tau, 0)$ is the operator $L$ in the Heisenberg picture.
\end{definition}
The physical significance of the $N-$point Green's function is that it determines the projection of the environment state on the $N-$particle subspace, as is made explicit in the following lemma (proved in the supplement \cite{supp}).
\begin{lemma} Let $\ket{\psi(t)} = U(t, 0) \big(\ket{\sigma}\otimes \ket{\textnormal{vac}}\big)$, where $\ket{\sigma}$ is a system state, then
\[
\ket{\psi(t)} = \sum_{N = 0}^\infty \ket{\psi^N(t)},
\]
where
\[
\ket{\psi^N(t)} =   \frac{1}{N!} \int_{\omega \in \mathbb{R}} F_N(\omega; t) \bigg[\prod_{i=1}^N v^*(\omega_i) a_{\omega_i}^\dagger \bigg]\ket{\sigma}\otimes \ket{\textnormal{vac}} d \omega,
\]
and $\forall \omega \in \mathbb{R}^N$,
\begin{align*}
F_N(\omega; t) = \int_{s\in [0, t]^N} G_N(s; t) e^{-i\omega\cdot(t - s)}d s.
\end{align*}
\end{lemma}
Furthermore, as is made precise in lemma  (proved in supplement \cite{supp}), the $N-$point Green's function is bounded, and consequently the norm of the $N-$particle projection of the environment state is also bounded.
\begin{lemma} If $v \in \mathbb{C}_b^\infty(\mathbb{R})$, then
\[\norm{G_N(s, t)} \leq \norm{L}^N \forall N \in \mathbb{N}, t \geq 0, s \in [0, t]^N, \]
and
\[
\norm{\ket{\psi^N(t)}}^2 \leq \frac{\norm{v}_\infty^{2N} \norm{L}^{2N} (2\pi t)^N}{N!} \ \forall N \in \mathbb{N}, t \geq 0.
\]
\end{lemma}
\edit{Our second assumption can be interpreted as a bound on the rate at which the local system can emit or absorb an excitation from the environment. Any physically reasonable model of the environment, despite its non-negligible couplings with high frequency environment modes, is expected to satisfy this assumption. }

\begin{assumption} $\forall t\geq 0, s \in [0, t]^{N - 1}$, the map $G_N(\{\cdot, s\}; t): [0, t] \to \mathfrak{L}(\mathcal{H}_S)$ is absolutely continuous and $\exists \gamma(t) > 0$ such that
\begin{align*}
\underset{\{s_0, s\} \in [0, t]^N}{\textnormal{esssup}}\norm{\partial_{s_0} G_N(\{s_0, s\}; t)} \leq  \gamma(t) \norm{L}^N.
\end{align*}
\end{assumption}
This assumption can be proved for two cases:
\begin{enumerate}
\item \emph{Markovian environments}, i.e.~environments with a frequency independent coupling constant ($v(\omega) = v_0$). In this case, an application of the quantum regression theorem can be used to show that assumption 2 is satisfied for such environments (proposition 3 in the supplement \cite{supp}).
\item \emph{Environments with a square integrable coupling constant,} assumption 2 can again be rigorously proven (proposition 4 in the supplement \cite{supp}).
\end{enumerate}

With these two assumptions, we can now prove the convergence of the pseudomode theory \cite{tamascelli2018nonperturbative} and star-to-chain transformation \cite{chin2010exact} for simulating non-Markovian quantum systems. Our first result, proved in the supplement \cite{supp}, rigorously shows that in a finite amount of time, the localized system cannot excite arbitrarily high frequencies in the environment, and the environment can thus be approximated within a finite energy window. More precisely,
\begin{theorem}\label{thm:cutoff} Suppose $v \in \textnormal{C}_b^\infty(\mathbb{R}) \cap \mathcal{S}'(\mathbb{R})$ is a coupling function such that assumptions 1 and 2 are satisfied. Denoting by $\rho(t)$ the reduced density matrix of the local system at time $t$ when an initial state $\ket{\sigma}\otimes\ket{\textnormal{vac}}$ is evolved under the Hamiltonian in Eq.~\ref{eq:hamiltonian} and by ${\rho}_{\omega_c}(t)$ the reduced density matrix of the local system at time $t$ when the same initial state is evolved under the Hamiltonian
\begin{align}\label{eq:hamiltonian_with_cutoff}
H_{\omega_c}(t) =& H_S(t) + \int_{-\infty}^{\infty} \omega a^\dagger_\omega a_\omega d\omega + \nonumber\\
&\int_{-\omega_c}^{\omega_c} \bigg( v(\omega) a_\omega L^\dagger + v^*(\omega) a^\dagger_\omega L\bigg)d\omega,
\end{align}
then
\begin{align*}
\norm{\rho(t) - {\rho}_{\omega_c}(t)}_\textnormal{tr} \leq \varepsilon(\omega_c, t),
\end{align*}
where $\varepsilon(\omega_c, t)$ is the cutoff error given by
\begin{align}
\varepsilon(\omega_c, t) = \frac{f_1(t)}{\sqrt{\omega_c}} + \int_0^t f_2(\tau) \sqrt{V(\omega_c, \tau)}d\tau.
\end{align}
Here $V(\omega_c, t)$ is defined in assumption 1 and
\begin{align*}
&f_1(\tau)= \sqrt{2}\norm{v}_\infty\norm{L} ( 2+ \gamma(\tau) \tau) e^{\pi\norm{v}_\infty^2 \norm{L}^2\tau}, \\
&f_2(\tau) = \sqrt{2}\norm{L}^2 \tau (1 + \gamma(\tau) \tau) e^{\pi\norm{v}_\infty^2 \norm{L}^2 \tau },
\end{align*}
where $\gamma(t)$ is introduced in assumption 2.
\end{theorem}
\edit{The dependence of the cutoff error on $\omega_c$ is a sum of two terms --- one that falls off as $O(1 / \sqrt{\omega_c})$, which can be interpreted as the consequence of introducing a rectangular frequency window on the emitted photon wave-packet, and a second term which depends on the error introduced in approximating the time-domain kernel $K_v$ within a frequency window (assumption 1). The magnitude of this error, as given by the functions $f_1, f_2$, depends on the strength of the coupling between the system and environment ($\propto\norm{v}_\infty,  \norm{L}$), as well as on the rate at which particles are exchanged with the environment ($\gamma(t)$ introduced in assumption 2). The result of theorem 1 is thus in-line with our intuition that a larger cutoff frequency is needed for systems which strongly couple to the environment, and absorb or emit particles very rapidly from the environment.} Furthermore, the error grows exponentially with time $t$ --- this is a consequence of the fact that the local system can in principle emit an arbitrarily large number of particles into the environment while only being constrained by the bound in lemma 2. In practice, we expect the errors grow only polynomially with $t$, with the polynomial to depend on the maximum number of particles that the local system can emit into the environment.

Next, we consider the pseudo-mode method \cite{tamascelli2018nonperturbative}, which approximates the non-Markovian dynamics of the local system by the Markovian dynamics of a larger system.
\begin{definition}[Pseudo-mode description]
\label{def:pseudomode}
An environment described by $M-$pseudomodes with parameters $\{(\omega_i, g_i, \kappa_i \geq 0) : i \in \{1, 2 \dots M\}\}$ has an associated Hilbert space $\mathcal{H}_\textnormal{aux}$ of $M$ bosonic modes with annihilation operators $a_0, a_1 \dots a_{M - 1}$. For a local system with Hilbert space $\mathcal{H}_S = \mathbb{C}^d$, time-dependent Hamiltonian $H_S(t) \in \mathfrak{L}(\mathbb{C}^d)$, interacting with the environment through the operator $L \in \mathfrak{L}(\mathbb{C}^d)$ with the initial system-environment state being $\ket{\sigma}\otimes \ket{\textnormal{vac}}$, its reduced state at time $t$ is given by $\rho(t) = \textnormal{Tr}_\textnormal{aux}(R(t))$, where $R(t)$ satisfies
\begin{align*}
&\dot{R}(t) = i[\hat{H}(t), R(t)] +\nonumber\\
&\qquad \sum_{i=0}^{M - 1} \frac{\kappa_i}{2}\big(2a_i R(t) a_i^\dagger - \{a_i^\dagger a_i, R(t)\}\big),
\end{align*}
with
\begin{align*}
&\hat{H}(t) = \nonumber\\
&\qquad H_S(t) + \sum_{i=0}^{M - 1} \omega_i a_i^\dagger a_i + \sum_{i=1}^M g_i(a_i L^\dagger + a_i^\dagger L),
\end{align*}
and $R(0) = \ket{\sigma}\bra{\sigma} \otimes (\ket{0}\bra{0})^{\otimes M}$.
\end{definition}
An environment described by $M-$pseudomodes with parameters $\{(\omega_i, g_i, \kappa_i \geq 0) : i \in \{1, 2 \dots M\}\}$ corresponds to a coupling function $\hat{v}$ which satisfies \cite{tamascelli2018nonperturbative}
\[
|\hat{v}(\omega)|^2 = \sum_{i=1}^M \frac{\kappa_i}{2\pi}\frac{g_i^2}{(\omega - \omega_i)^2 + \kappa_i^2 / 4}.
\]
In practice, to obtain a pseudomode description that approximates a given coupling function $v$, $|v|^2$ is approximated by a sum of lorentzians within a sufficiently large but finite frequency window, with each Lorentzian corresponding to an independent pseudomode. Our next result, proved in supplement \cite{supp}, shows that this procedure is guaranteed to converge. The convergence rate of the pseudomode approximation will, in general, depend on the details of the coupling function $v$ --- in particular, on the growth of its derivative $v'(\omega)$ with $\omega$, as well as on the falloff of the cutoff error $\varepsilon(\omega_c, t)$, introduced in theorem \ref{thm:cutoff}, with the cutoff frequency. For typical coupling functions, $|v'(\omega)| = O(\textnormal{poly}(\omega))$ and $\varepsilon(\omega_c, t) = O(\exp(O(t))\textnormal{poly}(\omega_c^{-1}))$. Under these assumptions, we show that the error incurred in the pseudomode approximation decreases polynomially with the number of pseudomodes.
\begin{theorem}[Pseudomode approximation]
Suppose $v \in \textnormal{C}^\infty_b(\mathbb{R}) \cap \mathcal{S}'(\mathbb{R})$ is a coupling function such that assumptions 1 and 2 are satisfied and let $\rho(t)$ be the reduced state of the local system after evolving an initial state $\ket{\sigma}\otimes \ket{\textnormal{vac}}$ using the Hamiltonian in Eq.~\ref{eq:hamiltonian}. Then, there exists a pseudomode description of the environment (definition \ref{def:pseudomode}) with $M$ bosonic modes which provides an approximation $\hat{\rho}(t)$ to the reduced state of the local system such that $\norm{\rho(t) - \hat{\rho}(t)}_\textnormal{tr} \to 0$ as $M \to \infty$. Furthermore, if $|v'(\omega)| = O(\textnormal{poly}(\omega))$ and the cutoff error $\varepsilon(\omega_c, t) = O(\exp(O(t))\textnormal{poly}(\omega_c^{-1}))$, then there exists a pseudomode description of the non-Markovian system with $M$ bosonic modes such that the trace-norm error in approximating the reduced local system state at time $t$ scales as $O(\exp(O(t))\textnormal{poly}(M^{-1}))$.
\end{theorem}

Finally, we consider the star-to-chain transformation, which maps the non-Markovian environment to a 1D bosonic tight-binding model with the local system effectively coupled to the first bosonic mode.
\begin{definition}[Chain description]
\label{def:chain_descr}
An environment described by a chain of $M$ bosonic modes with parameters $\{(\omega_i, g_i) : i \in \{0, 1, 2, 3 \dots M - 1\}\}$ has an associated Hilbert space $\mathcal{H}_\textnormal{aux}$ of $M$ bosonic modes with annihilation operators $a_0, a_1 \dots a_{M - 1}$,
For a local system with Hilbert space $\mathcal{H}_S = \mathbb{C}^d$, time-dependent Hamiltonian $H_S(t) \in \mathfrak{L}(\mathbb{C}^d)$ interacting with the environment through the operator $L \in \mathfrak{L}(\mathbb{C}^d)$ with the initial system-environment state being $\ket{\sigma}\otimes \ket{\textnormal{vac}}$, its reduced state at time $t$ is given by $\rho(t) = \textnormal{Tr}_\textnormal{aux}(\hat{U}(t, 0)\ket{\sigma}\bra{\sigma} \otimes (\ket{0}\bra{0})^{\otimes M}\hat{U}(0, t))$, where $\hat{U}(\tau_1, \tau_2)$ is the propagator corresponding to the Hamiltonian
\[
\hat{H}(t) = H_S(t) + g_0 (L^\dagger a_0 + a_0^\dagger L) + H_E,
\]
with
\[
H_E = \sum_{i = 0}^{M - 1}\omega_i a_i^\dagger a_i + \sum_{i = 1}^{M-2}g_i(a_i a_{i + 1}^\dagger + a_{i+1}a_i^\dagger).
\]
\end{definition}
A chain transformation of the environment can be explicitly constructed by using the Lanczos iteration --- this proceeds by first introducing a frequency cutoff $\omega_c$, and then starting from the mode $a_0 \propto \int_{-\omega_c}^{\omega_c} v(\omega)a_\omega d\omega$, applying the Lanczos iteration with respect to the environment Hamiltonian. This yields the parameters $\{(\omega_i, g_i) : i \in \{0, 1, 2 \dots M - 1\}\}$ for the chain description of the environment. By exploiting the previously established connection between the star-to-chain transformation and orthogonal polynomials \cite{chin2010exact, de2015discretize}, we show (supplement \cite{supp}) that, under assumptions 1 and 2, the star-to-chain transformation will converge for sufficiently large number of modes and that the convergence rate is polynomial in the number of modes.
\begin{theorem}[Star-to-chain transformation] 
Suppose $v \in \textnormal{C}^\infty_b(\mathbb{R}) \cap \mathcal{S}'(\mathbb{R})$ is a coupling function such that assumptions 1 and 2 are satisfied and let $\rho(t)$ be the reduced state of the local system after evolving an initial state $\ket{\sigma}\otimes \ket{\textnormal{vac}}$ using the Hamiltonian in Eq.~\ref{eq:hamiltonian}. Then, there exists a chain description of the environment with $M$ modes (definition \ref{def:chain_descr}) that provides an approximation $\hat{\rho}(t)$ to the reduced state of the local system such that $\norm{\rho(t) - \hat{\rho}(t)}_\textnormal{tr} \to 0$ as $M \to \infty$. Furthermore, if the cutoff error $\varepsilon(\omega_c, t) = O(\exp(O(t))\textnormal{poly}(\omega_c^{-1}))$, then the trace-norm error in approximating the reduced local system state at time $t$ scales as $O(\exp(O(t))\textnormal{poly}(M^{-1}))$.
\end{theorem} \ \\

In conclusion, our work provides a rigorous analysis of Markovian dilations to non-Markovian open quantum systems. We show that the finite-time dynamics of a wide class of non-Markovian quantum systems can always be well approximated by a larger Markovian system, and we also provide theoretical scalings of how much larger the effective Markovian system is. Several questions of interest to open quantum system theory are left open in this work. One of the questions that we leave unanswered is to study the class of coupling functions, with possibly distributional kernels, for which the resulting system-environment dynamics is well defined. A rigorous study of this problem would be relevant to advancing the mathematical understanding of non-Markovian open quantum system models, and could lead to a general proof of assumption 2. Another direction to pursue would be to improve the exponential dependence of the error estimates on time $t$, or to identify settings in which these error estimates are tight. \nocite{trivedi2018few, caneva2015quantum, mcclarren2018gauss}
\begin{acknowledgements}
We thank Martin Plenio, Mischa Woods, Christiane Koch and Archak Purkayastha for pointing us to relevant literature on this topic. RT acknowledges Max Planck Harvard research center for quantum optics (MPHQ) postdoctoral fellowship. We  acknowledge  support  from the  ERC  Advanced  Grant  QUENOCOBA  under  the EU  Horizon  2020  program  (grant  agreement  742102) and  from  the  Deutsche  Forschungsgemeinschaft  (DFG, German Research Foundation) under the project number414325145 in the framework of the Austrian Science Fund(FWF): SFB F7104.
\end{acknowledgements}

\twocolumngrid
\bibliography{references.bib}
\end{document}


\title{Supplement to convergence for Markovian dilations to non-Markovian bosonic gaussian environments}
\author{Rahul Trivedi$^{1, 2}$}
\email{rahul.trivedi@mpq.mpg.de}
\author{Daniel Malz$^{1, 2}$}
\author{J.~Ignacio Cirac$^{1, 2}$}
\address{$^1$Max-Planck-Institut für Quantenoptik, Hans-Kopfermann-Str.~1, 85748 Garching, Germany.\\
$^2$Munich Center for Quantum Science and Technology (MCQST), Schellingstr. 4, D-80799 Munich, Germany.}

\date{\today}

\maketitle

{
\tableofcontents
\makeatletter
\let\toc@pre\relax
\let\toc@post\relax
\makeatother 
}
\section{Notation and mathematical prelimnaries}
\noindent\emph{Function spaces and analysis}: For $v \in \mathbb{R}^N$ and $u \in \mathbb{R}^M$, we will denote by $\{v, u\}$ a vector in $\mathbb{R}^{N + M}$ whose first $N$ entries are elements of $v$ and the next $M$ entries are elements of $u$.

All the integrals in this paper are Lebesgue integrals and with respect to the Lebesgue measure over $\mathbb{R}^N$. For a measure space $(\Omega, \mathfrak{F}, \mu)$, given a measurable map $f:\Omega \to \mathbb{R}$, we define its essential supremum via $\text{esssup}_{x\in \Omega}f(x) = \inf \{a \in \mathbb{R}\ |\ \mu(f^{-1}([a, \infty])) = 0 \}$. The essential supremum can be interpreted as the supremum of a function obtained on ignoring subsets of its domain with zero measure. A propositional function $\textnormal{P}:\Omega \to \{\text{True}, \text{False}\}$ is said to be True \emph{almost everywhere} in $\Omega$ if the $\mu\big(\{x\ |\ \textnormal{P}(x) \text{ is False}\} \big)= 0$. This will be denoted as $\forall x \in_{a.e.}\Omega: \ \text{P}(x) \text{ is True}$.

We will denote by $\text{L}^2(\mathbb{R}^D)$ the set of complex-valued square integrable functions over $\mathbb{R}^D$. We will denote by $\text{B}(\mathbb{R}^D)$ the set of bounded functions over $\mathbb{R}^D$. For $f \in \text{B}(\mathbb{R}^D)$, we denote by $\norm{f}_\infty$ the maximum magnitude of the function i.e. $\norm{f}_\infty = \sup_{x \in \mathbb{R}^D} 
|f(x)|$. For a closed interval $I \subset \mathbb{R}$, we will denote by $\text{C}^k(I)$ the set of functions $f:I \to \mathbb{C}$ which are $k-$continuously differentiable. In particular, $\text{C}^0(I)$ is the set of continuous complex-valued functions with domain $I$ and $\text{C}^\infty(I)$ is the space of smooth complex-valued functions with domain $I$. We will also denote by $\textrm{C}^k_b(I)$ the space of bounded $k-$times continuously differentiable functions --- on compact intervals $I$, this will coincide with $\textrm{C}^k(I)$. We will denote the space of tempered distributions over $\mathbb{R}$ by $\mathcal{S}'(\mathbb{R})$ --- $\textrm{C}^\infty_b(\mathbb{R})\cap \mathcal{S}'(\mathbb{R})$ is the space of tempered distributions that correspond to multiplication with bounded and smooth functions. Finally, for any $\alpha>0$, $\text{rect}_\alpha:\mathbb{R}\to \mathbb{R}$ to be denote the rectangular windowing function defined by
\[
\text{rect}_\alpha(x) = \begin{cases}
1 & \text{ if } |x| \leq \alpha\\
0 & \text{ if } |x| > \alpha.
\end{cases}
\]

For $a < b$, we will denote by $\text{AC}([a, b])$ the space of all functions $f:[a, b] \to \mathbb{C}$ that are absolutely continuous. We recall that $f \in \text{C}^0([a, b])$ is absolutely continuous if $\exists g: [a, b] \to \mathbb{C}$ such that $g$ is integrable and
\[
\forall \ x \in [a, b]: \ f(x) = f(a) + \int_a^x g(x')dx'.
\]
An immediate consequence of this definition is that the function $f$ has a derivative, which we denote by $\partial_x f $, almost everywhere on $[a, b]$. Furthermore, we can define a norm  $\norm{\cdot}^{\mathcal{S}}_{[a, b]}:\text{AC}([a, b])\to \mathbb{R}^+$ such that $\forall  f \in \text{AC}([a, b])$
\begin{align}\label{eq:seminorm_sv}
\norm{f}^{\mathcal{S}}_{[a, b]} = \sup_{x \in [a, b]} |f(x)| + (b - a) \underset{x \in [a, b]}{\text{ esssup }} \big|\partial_x f(x)\big|.
\end{align}
A map $f: [a, b]^2 \to \mathbb{C}$ will be called symmetric if $\forall s_1, s_2 \in [a, b]: \ f(s_1, s_2) = f^*(s_2, s_1)$.  We will denote by $\text{AC}_\text{sym}([a, b]^2)$ the space of symmetric, continuous functions $f:[a, b]^2 \to \mathbb{C}$ which satisfy the following two conditions
\begin{enumerate}
\item $f$ is absolutely continuous in either of its arguments i.e. $\forall y \in [a, b],$ the map $\varphi_y: [a, b] \to \mathbb{C}$ defined by $\varphi_y(x) = f(x, y)$ is absolutely continuous.
\item For $y \in_{a.e.}[a, b]$, $\partial_y f(x, y)$ exists $\forall x \in [a, b]$ and the map $\varphi_y^d: [a, b] \to \mathbb{C}$ defined by $\varphi_y^d(x) = \partial_y f(x, y)$ is absolutely continuous.
\end{enumerate}
We note that the partial derivatives of $f$ with respect to either and both of its arguments, $\partial_x f$, $\partial_y f$, $\partial^2_{x,y} f$, exists almost everywhere on $[a, b]^2$. Finally, we define a norm $\norm{\cdot}_{[a, b]^2}^{\mathcal{S}}: \text{AC}_\text{sym}([a, b]^2) \to \mathbb{R}^+$ such that $\forall  f \in \text{AC}_\text{sym}([a, b]^2)$
\begin{align}\label{eq:seminorm_dv}
\norm{f}^{\mathcal{S}}_{[a, b]^2} &=  \sup_{x, y \in [a, b]} |f(x, y)| + \nonumber \\ &(b - a) \bigg(\underset{x, y \in [a, b]}{\text{esssup}} \big|\partial_x f(x, y)\big| + \underset{{x, y \in [a, b]}}{\text{esssup}} \big|\partial_y f(x, y)\big|\bigg) + \nonumber\\ &(b-a)^2\underset{{x, y \in [a, b]}}{\text{esssup}} \big|\partial^2_{x, y} f(x, y)\big| .
\end{align}
A symmetric map $K$ is also positive if
\[
\forall f \in \text{C}^0([a, b]): \ \int_{s_1, s_2 = a}^b K(s_1, s_2) f^*(s_1) f(s_2) ds_1 ds_2 \geq 0.
\]
We define $\text{AC}_\text{sym}^{\succeq 0}([a, b]^2) = \{f \in \text{AC}_\text{sym}([a, b]^2) | f \text{ is positive.}\}$

\ \\
\noindent\emph{Norms}: For a Hilbert space $\mathcal{H}$, the Banach space of bounded operators over $\mathcal{H}$ will be denoted by $\mathfrak{L}(\mathcal{H})$. For elements of a Hilbert space $\mathcal{H}$, $\ket{\psi} \in \mathcal{H}$, $\norm{\ket{\psi}}$ will denote the norm induced by $\mathcal{H}'$s inner product. For a bounded operator $O:\mathcal{H} \to \mathcal{H}$, we will denote by $\norm{O}$ the operator norm induced by the norm on $\mathcal{H}$ i.e.~$\norm{O} = \sup_{\ket{\psi} \in \mathcal{H}} \norm{O\ket{\psi}} / \norm{\ket{\psi}}$. For a bounded and trace-class operator $O: \mathcal{H} \to \mathcal{H}$, its trace-norm is given by $\norm{O}_\text{tr} = \text{tr}[\sqrt{O^\dagger O}]$. The trace-norm will be used to quantify distance between two density operators. It is useful to note that given two states $\ket{\psi_1}, \ket{\psi_2} \in \mathcal{H}$ with unit norm, then the operator $\ket{\psi_1}\bra{\psi_1} - \ket{\psi_2}\bra{\psi_2}$ is bounded and trace-class and its trace-norm satisfies
 \begin{align}\label{eq:tr_norm_l2}
 \norm{\ket{\psi_1}\bra{\psi_1} - \ket{\psi_2}\bra{\psi_2}}_\text{tr} \leq \sqrt{2}\norm{\ket{\psi_1} - \ket{\psi_2}}.
 \end{align}
 This property, together with the contractivity of the trace-norm under partial trace, allows us to translate an error in between two states of a composite quantum system to an error between the density operators of a subsystem of the composite quantum system.
\ \\ \ \\
\noindent\emph{Error analysis}: Given two functions $f, g:[0, \infty) \to [0, \infty)$ of parameter $h$ then the notation $f = O(g) \text{ as } h \to h_0 \implies \exists c>0:  f(h) \leq cg(h)$ as $h \to h_0$. The notation $f = \Theta(g)$ as $h \to h_0 
 \implies \exists c_-, c_+ > 0: c_- g(h) \leq f(h) \leq c_+g(h)$ as $h \to h_0$. Furthermore, $f = O(g_1) + O(g_2)$ as $h \to h_0$ if $\exists f_1, f_2$ such that $f = f_1 + f_2$ and $f_1 = O(g_1)$ as $h \to h_0$ and $f_2 = O(g_2)$ as $h \to h_0$.

\section{$N$-point Green's function and the environment state}\label{app:gfunc}
\edit{\noindent In this section, we relate the $N-$point Green's function introduced in the main text to the joint state of the local system and environment (lemma 1) and also provide a short proof of lemma 2 from the main text. For generality, in this section we consider the case where the environment is in an initially excited state $\ket{\textrm{E}} \in \mathcal{H}_E$ --- the results of the main text correspond to the setting where $\ket{\textrm{E}} \equiv \ket{\text{vac}}$, while in section \ref{sec:initial_env}, we consider an initially excited environment state. \\ \ \\

\noindent\textbf{Definition 2 (Restated, for initially excited environment)} \emph{ Let $\ket{\textnormal{E}} \in \mathcal{H}_E$ be an environment state, then $\forall s \in [0, t]^N$, $G_N^\textnormal{E}(s; t) \in \mathfrak{L}(\mathcal{H}_S)$ via
\[
G_N^{\ket{\textnormal{E}}}(s; t) = \bra{\textnormal{vac}} U(t, 0) \mathcal{T}\bigg[\prod_{i=1}^N L(s_i) \bigg]\ket{\textnormal{E}}
\]
where $U(\tau_1, \tau_2)$ is the propagator for the joint system-environment dynamics and $L(\tau) = U(0, \tau) L U(\tau, 0)$ is the operator $L$ in the Heisenberg picture.} \\ \ \\

\noindent\textbf{Lemma 1 (Restated, for initially excited environment)} \emph{Let $\ket{\psi(t)} = U(t, 0)(\ket{\sigma} \otimes \ket{\textrm{E}})$ where $\ket{\sigma} \in \mathcal{H}_S$ is an initial system state and $\ket{\textrm{E}} \in \mathcal{H}_E$ is the initial environment state, then
\[
\ket{\psi(t)} = \sum_{N = 0}^\infty \ket{\psi^N(t)},
\]
where
\[
\ket{\psi^N(t)} = \frac{1}{N!} \int_{\omega \in \mathbb{R}} F_N(\omega; t) \bigg[ \prod_{i=1}^N v^*(\omega_i)a_{\omega_i}^\dagger \bigg] \ket{\sigma}\otimes \ket{\textnormal{vac}} d\omega,
\]
and $\forall \omega \in \mathbb{R}^N$,
\[
F_N(\omega; t) = \int_{s \in [0, t]^N} G_N(s; t) e^{-i\omega \cdot (t - s)} ds.
\]
}
\noindent\emph{Proof}:
The state $\ket{\psi(t)}$ admits a representation of the form
\begin{align}\label{eq:expansion_psi_t}
\ket{\psi(t)} = \sum_{N = 0}^\infty \ket{\psi^N(t)} = \sum_{N=0}^\infty \sum_{i = 1}^d \frac{1}{N!} \int_{\omega\in\mathbb{R}^N}\psi^N_i(\omega; t) \bigg(\prod_{j=1}^N a_{\omega_j}^\dagger\bigg) \ket{e_i}\otimes \ket{\text{vac}},
\end{align}
where $\{\ket{e_i}, i \in \{1, 2 \dots d\}\}$ is a basis for the $d-$dimensional local system and
\[
\psi^N_i(\omega; t) = \bra{e_i} \otimes \bra{\text{vac}} \bigg(\prod_{i = 1}^M a_{\omega_i}\bigg) U(t, 0)\ket{\sigma}\otimes \ket{\text{vac}} = \bra{e_i} \otimes \bra{\text{vac}} U(t, 0)\mathcal{T}\bigg(\prod_{i = 1}^M a_{\omega_i}(t)\bigg)\ket{\sigma}\otimes \ket{\text{vac}},
\]
where $a_\omega(t) = U(0, t) a_\omega U(t, 0)$ and the time-ordering operator can be trivially introduced since all the operators are evaluated at the same time. From Heisenberg equations of motion, it immediately follows that
\[
a_\omega(t) = a_\omega(0) e^{-i\omega t} - iv^*(\omega) \int_0^t L(s) e^{-i\omega(t-s)}ds,
\]
from which it immediately follows that
\begin{align}\label{eq:comp_explicit}
\psi^N_i(\omega; t) = (-i)^N \bigg(\prod_{j =1}^N v^*(\omega_j)\bigg)\bra{e_i}\otimes \bra{\text{vac}}F_N(\omega; t)\ket{\sigma}\otimes \ket{\text{vac}},
\end{align}
Substituting Eq.~\ref{eq:comp_explicit} into Eq.~\ref{eq:expansion_psi_t} and using $\sum_{i=1}^d \ket{e_i}\bra{e_i} = \text{I}$, we obtain that
\begin{align}\label{eq:N_photon_component}
\ket{\psi^N(t)} =   \frac{1}{N!} \int_{\omega \in \mathbb{R}} F_N(\omega; t) \bigg[\prod_{i=1}^N v^*(\omega_i) a_{\omega_i}^\dagger \bigg]\ket{\sigma}\otimes \ket{\text{vac}} d \omega.
\end{align}

\noindent\textbf{Lemma 2 (Restated, for initially excited environment)} \emph{$\forall \ket{\textrm{E}}\in\mathcal{H}_E, N \in \mathbb{N}, t \geq 0, s \in [0, t]^N$, it follows that
\[
\norm{G_N^{\ket{\textnormal{E}}}(s; t)} \leq \norm{L}^N,
\]
and
\[
\norm{\ket{\psi^N(t)}}^2 \leq \frac{\norm{v}_\infty^{2N}\norm{L}^{2N} (2\pi t)^N}{N!}.
\]
}
\emph{Proof}: This follows from noting that $\forall \ket{\textrm{E}}\in\mathcal{H}_E, N \in \mathbb{N}, t \geq 0, s \in [0, t]^N$
\[
\norm{U(t, 0) \mathcal{T}\bigg[\prod_{i=1}^N L(s_i)\bigg]} \leq \norm{L}^N.
\]
Next, we note from Eq.~\ref{eq:N_photon_component} that
\begin{align*}
\norm{\ket{\psi^N(t)}}^2 &= \frac{1}{N!}\int_{\omega \in \mathbb{R}^N} \bra{\sigma}F_N^\dagger(\omega; t) F_N(\omega; t) \ket{\sigma} \bigg(\prod_{i=1}^{N}|v(\omega_i)|^2\bigg) d\omega \\
&\leq \frac{\norm{v}_\infty^{2N}}{N!} \int_{\omega \in \mathbb{R}^N}\bra{\sigma} F_N^\dagger(\omega; t) F_N(\omega; t) \ket{\sigma} \nonumber\\
&= \frac{\norm{v}_\infty^{2N} (2\pi)^N}{N!} \int_{s \in [0, t]^N} \bra{\sigma} G_N^\dagger(s; t) G_N(s; t) \ket{\sigma}ds \nonumber\\
& \leq \frac{\norm{v}_\infty^{2N} (2\pi)^N}{N!} \int_{s\in [0, t]^N} \norm{L}^{2N} dt \nonumber\\
&= \frac{\norm{v}_\infty^{2N}\norm{L}^{2N} (2\pi t)^N}{N!}.
\end{align*}
This finishes the proof of the lemma. $\square$.}

\section{Settings where assumptions 1 and 2 are provable}
\subsection{Assumption 1}\label{app:assump_1}
\edit{For completeness, we restate assumption 1 from the main text, and then prove it for several classes of coupling functions encountered in practice.
\begin{definition} \textnormal{\textbf{(Restated)}} Let $v \in \mathbb{C}_b^\infty(\mathbb{R}) \cap \mathcal{S}'(\mathbb{R})$, and let $K_v \in \mathcal{S}'(\mathbb{R})$ be the corresponding kernel defined by
\[
K_v(t) = \int_{-\infty}^\infty  |v(\omega)|^2e^{-i\omega t} d\omega,
\] 
then for $t \geq 0$ define a quasi-norm $\ell(K_v, t)$ of $K_v$ 
\[
\ell(K_v, t) = \sup_{f \in \text{AC}_\textnormal{sym}^{\succeq 0}([0, t]^2)} \frac{1}{\norm{f}_{[0, t]^2}^\mathcal{S}} \int_{s_1, s_2 = 0}^t K_v(t - s_1) K^*(t - s_2) f(s_1, s_2) ds_1 ds_2.
\]
\end{definition}
\noindent \textbf{Assumption 1 (Restated)}: \emph{The coupling function $v \in \mathbb{C}_b^\infty(\mathbb{R}) \cap \mathcal{S}'(\mathbb{R})$ is such that there is a function $V(\omega_c, t)$ which vanishes as $\omega_c \to \infty \ \forall t \geq 0$ and
\[
\ell(K_v - K_{v_{\omega_c}}) \leq V(\omega_c, t).
\]
where $v_{\omega_c}(\omega) = v(\omega)$ if $|\omega| \leq \omega_c$ and otherwise 0.
}
}
\begin{proposition}\label{prop:sq_int_assump_1} Assumption 1 is true for coupling functions $v$ that are bounded and square integrable, $v \in \textnormal{B}(\mathbb{R}) \cap \textnormal{L}^2(\mathbb{R})$.
\end{proposition}
\emph{Proof}: We consider constants $v:\mathbb{R}\to \mathbb{C}$ that are smooth, bounded and square integrable. We note that in this case, for a symmetric and positive kernel $K: [0, t]^2 \to \mathbb{C}$,
\begin{align*}
&\lim_{\Omega \to \infty}\int_{\omega_c\leq |\omega_1|, |\omega_2| \leq \Omega} |v(\omega_1)|^2 |v(\omega_2)|^2 \bigg(\int_{s_1, s_2 = 0}^t K(s_1, s_2) e^{i\omega_2(t - s_2)}e^{-i\omega_1(t - s_1)} ds \bigg)d\omega_1 d\omega_2   = \nonumber \\
&\bigg| \int_{\omega_c\leq |\omega_1|, |\omega_2|} |v(\omega_1)|^2 |v(\omega_2)|^2 \bigg(\int_{s_1, s_2 = 0}^t K(s_1, s_2) e^{i\omega_2(t - s_2)}e^{-i\omega_1(t - s_1)} ds \bigg)d\omega_1 d\omega_2 \bigg | \leq \nonumber \\
&\int_{\omega_c\leq |\omega_1|, |\omega_2|} |v(\omega_1)|^2 |v(\omega_2)|^2 \bigg(\int_{s_1, s_2 = 0}^t |K(s_1, s_2) |ds_1 ds_2 \bigg)d\omega_1 d\omega_2  \leq t^2\bigg(\int_{\omega_c \leq |\omega|}|v(\omega)|^2 d\omega\bigg)^2\sup_{s_1, s_2 \in [0, t]} |K(s_1, s_2)|.
\end{align*}
By noting that $\sup_{s_1, s_2 \in [0, t]} |K(s_1, s_2)| \leq \norm{K}_{[0, t]^2}^\mathcal{S}$, it follows that square-integrable coupling constants satisfy the conditions of assumption 1 with $V(\omega_c, t)$ being given by
\[
V(\omega_c, t) = t^2 \int_{|\omega| \geq \omega_c} |v(\omega)|^2 d\omega \ \qquad \qquad \qquad \qquad\square
\]

We note that the case of a square integrable coupling constant contains the physically important case of a lorentzian coupling function, which often arises in studying the interaction of quantum systems with lossy resonators. Next, we show that assumption 1 is also true for environments whose coupling function is a sum of a discrete number of harmonic terms. Such coupling functions arise in models that consider a local quantum system with point couplings to a one-dimensional propagating field. We first provide a bound on the error incurred in approximating a delta function by a sinc function.
\begin{lemma}\label{lemma:endpt_appx_delta_fun} Let $f \in \textnormal{AC}([0, a])$, then $\forall p > 0$
\[
\bigg|\frac{\pi}{2} f(0) - \int_{0}^{a}\frac{\sin px}{x} f(x)dx\bigg| \leq O\bigg(\frac{1}{\sqrt{ap}}\bigg) \norm{f}^{\mathcal{S}}_{[0, a]}.
\]
\end{lemma}
\emph{Proof}: We begin by using the Dirchlet integral, $\int_0^\infty \sin x / x dx = \pi / 2$, from which it follows that
\[
\bigg|\frac{\pi}{2}f(0) - \int_0^a \frac{\sin px}{x} f(x) dx \bigg| \leq \bigg|\int_{0}^a \big(f(x) - f(0)\big) \frac{\sin px}{x}dx\bigg| + |f(0)|\bigg| \int_a^\infty \frac{\sin px}{x}dx\bigg|.
\]
The second term is easily bounded
\begin{align*}
 |f(0)|\bigg| \int_a^\infty \frac{\sin px}{x}dx\bigg| &\leq \bigg(\sup_{x \in [0, a]} |f(x)|\bigg) \bigg|-\frac{\cos pa}{pa} + \int_a^\infty \frac{\cos px}{px^2}dx \bigg| \nonumber \\
 &\leq \bigg(\sup_{x \in [0, a]} |f(x)|\bigg) \bigg(\frac{1}{pa} + \int_a^\infty \frac{1}{px^2}dx\bigg) = \bigg(\sup_{x \in [0, a]} |f(x)|\bigg) \frac{2}{pa}.
\end{align*}
To bound the first term, we split the integral at $b(p) \in [0, a]$, where $b(p)$, to be specified later, vanishes as $p\to \infty$:
\[
\bigg|\int_{0}^a \big(f(x) - f(0)\big) \frac{\sin px}{x}dx\bigg| \leq \int_{0}^{b(p)} \big|f(x) - f(0)\big| \frac{dx}{x} + \bigg|\int_{b(p)}^a \big(f(x) - f(0)\big) \frac{\sin px}{x}dx\bigg|.
\]
Noting that $|f(x) - f(0)| \leq |x|\ \text{esssup}_{x \in [0, a]} |f'(x)|$, we obtain
\[
\int_{0}^{b(p)} \big|f(x) - f(0)\big| \frac{dx}{x} \leq b(p)\  \underset{x \in [0, a]}{\text{ esssup }} |f'(x)|.
\]
Futhermore, using integration by parts, we obtain
\begin{align*}
\bigg| \int_{b(p)}^a \big(f(x) - f(0)\big) \frac{\sin px}{x}dx\bigg| &= \bigg|\frac{f(b(p)) \cos (pb(p))}{p b(p)} - \frac{f(a) \cos pa}{pa} + \frac{1}{p} \int_{b(p)}^a \bigg(\frac{f'(x)}{x} - \frac{f(x) - f(0)}{x^2}\bigg) \cos px dx\bigg|, \nonumber \\
&\leq \frac{3}{p}\bigg(\frac{1}{b(p)} + \frac{1}{a}\bigg) \sup_{x\in[0, a]} |f(x)| + \frac{\log(a / b(p))}{p}  \underset{x \in [0, a]}{\text{ esssup }}  |f'(x)|.
\end{align*}
Using these estimates and noting that $\sup_{x \in [0, a]} |f(x)| \leq \norm{f}_{[0, a]}^\mathcal{S}$ and $\text{esssup}_{x\in [0, a]}|f'(x)| \leq \norm{f}_{[0, a]}^\mathcal{S} / a$, we obtain
\[
\bigg|\frac{\pi}{2}f(0) - \int_0^a \frac{\sin px}{x} f(x) dx\bigg| \leq \bigg(\frac{b(p)}{a} + \frac{1}{p}\bigg(\frac{3}{b(p)} + \frac{5}{a}\bigg) + \frac{\log(a / b(p))}{ap}\bigg) \norm{f}_{[0, a]}^\mathcal{S}.
\]
Since this bound holds for any $b(p) \in [0, a]$, we choose $b(p) = \Theta(\sqrt{a / p})$ which proves the lemma $\square$.
\begin{lemma}\label{lemma:outside_delta} Let $f \in \textnormal{AC}([a, b])$ for $0 < a < b$, then $\forall p > 0$ 
\[
\bigg|\int_{a}^b \frac{\sin px}{x}f(x) dx\bigg| \leq O\bigg(\frac{\log(b / a)}{ap}\bigg)\norm{f}_{[a, b]}^\mathcal{S}
\] 
\end{lemma}
\emph{Proof}: This follows straightforwardly from integration by parts.
\[
\int_a^b \frac{\sin px}{x}f(x) dx = \frac{f(a) \cos pa}{pa} - \frac{f(b) \cos pb}{pb} + \frac{1}{p}\int_{a}^b \bigg(\frac{f'(x)}{x} - \frac{f(x)}{x^2}\bigg) \cos px \ dx,
\]
and consequently
\[
\bigg|\int_a^b \frac{\sin px}{x}f(x) dx\bigg| \leq  \bigg(\frac{2}{pa} + \frac{2}{pb} + \frac{\log(b / a)}{ap} \bigg)\norm{f}^\mathcal{S}_{[a, b]},
\]
from which the lemma statement follows $\square$.
\begin{lemma}\label{lemma:final_delta_fun_bound}Let $f \in \textnormal{AC}([a, b])$, then for $p > 0$ and $y \in \mathbb{R}$
\[
\bigg|\int_a^b \bigg(\pi\delta(x - y) - \frac{\sin p(x - y)}{x - y}\bigg) f(x)dx  \bigg| \leq \xi(a, b, y, p)\norm{f}_{[a, b]}^\mathcal{S}.
\]
where $\xi(a, b, y, p) \to 0$ as $p \to \infty$ and $\delta(\cdot)$ is the dirac-delta function, whose action on a continuous function $f$ within an interval $[a, b]$ is given by
\[
\int_a^b \delta(x - y) f(x) dx = \begin{cases}
f(y) & \text{if } y \in (a, b) \\
f(y) / 2 & \text{if } y \in \{a, b\} \\
0 & \text{otherwise}
\end{cases}.
\]
\end{lemma}
\emph{Proof}: We consider three different cases
\begin{enumerate}
\item If $y \in \{a, b\}$, then it follows from lemma \ref{lemma:endpt_appx_delta_fun} that
\[
\bigg| \int_a^b \bigg(\pi\delta(x - y) - \frac{\sin p(x - y)}{x - y}\bigg) f(x) dx \bigg| \leq O\bigg(\frac{1}{\sqrt{|a - b|p}}\bigg) \norm{f}_{[a, b]}^\mathcal{S}.
\]
\item If $y < a$ or $y > b$, then it follows from lemma \ref{lemma:outside_delta} that
\[
\bigg| \int_a^b \bigg(\pi\delta(x - y) - \frac{\sin p(x - y)}{x - y}\bigg) f(x) dx \bigg| \leq O\bigg(\frac{|\log(|y - a| / |y - b|)|}{p\min(|y - a|, |y - b|)}\bigg) \norm{f}_{[a, b]}^\mathcal{S}.
\]
\item If $y \in (a, b)$, then
\begin{align*}
&\bigg| \int_a^b \bigg(\pi\delta(x - y) - \frac{\sin p(x - y)}{x - y}\bigg) f(x) dx \bigg|  \nonumber \\ 
&\leq\bigg|\frac{\pi}{2}f(y) - \int_a^y \frac{\sin p(x - y)}{x - y} f(x) dx \bigg|  + \bigg| \frac{\pi}{2}f(y) - \int_y^b \frac{\sin p(x - y)}{x - y} f(x) dx \bigg| \nonumber \\
&\leq O\bigg(\frac{1}{\sqrt{|y - a|p}}\bigg) \norm{f}_{[a, y]}^\mathcal{S} + O\bigg(\frac{1}{\sqrt{|y - b|p}}\bigg)\norm{f}_{[y, b]}^\mathcal{S}\nonumber\\
&\leq O\bigg(\frac{1}{\sqrt{\min(|y - a|, |y - b|)p}}\bigg)\norm{f}_{[a, b]}^\mathcal{S}.
\end{align*}
Here we have used lemma \ref{lemma:endpt_appx_delta_fun} and the fact that $\forall y \in (a, b): \ \norm{f}_{[a, y]}^\mathcal{S} \leq \norm{f}_{[a, b]}^\mathcal{S}$ and $\norm{f}_{[y, b]}^\mathcal{S} \leq \norm{f}_{]a, b]}^\mathcal{S}$ $\square$.
\end{enumerate}
\begin{lemma}\label{lemma:delta_appx_kernel} For $K \in \textnormal{AC}_\textnormal{sym}([a, b]^2)$ and $y_1, y_2 \in \mathbb{R}$, then $\forall p > 0$
\begin{align*}
\bigg|\int_{x_1, x_2 = a}^b\bigg(\pi\delta(x_1 - y_1) - \frac{\sin p(x_1 - y_1)}{x_1 - y_1}\bigg)  \bigg(\pi\delta(x_2 - y_2) - \frac{\sin p(x_2 - y_2)}{x_2 - y_2}\bigg) K(x_1, x_2) dx_1 dx_2\bigg|\nonumber\\
 \leq \xi(a, b, y_1, p) \xi(a, b, y_2, p)\norm{K}_{[a, b]^2}^\mathcal{S}.
\end{align*}
\end{lemma}
\emph{Proof}: For $y \in [a, b]$, we define $\varphi_y:[a, b] \to \mathbb{C}$ via
\[
\varphi_y(x) = \int_a^b \bigg(\pi \delta(x' - y) - \frac{\sin p(x' - y)}{x' - y}\bigg) K(x, x') dx',
\]
such that
\begin{align*}
\int_{x_1, x_2 = a}^b \bigg(\pi\delta(x_1 - y_1) - \frac{\sin p(x_1 - y_1)}{x_1 - y_1}\bigg)  \bigg(\pi\delta(x_2 - y_2) - \frac{\sin p(x_2 - y_2)}{x_2 - y_2}\bigg) K(x_1, x_2) dx_1 dx_2 =\\ \int_a^b \bigg(\pi \delta(x_1 - y_1) - \frac{\sin p(x_1 - y_1)}{x_1 - y_1}\bigg) \varphi_{y_2}(x_1) dx_1.
\end{align*}
Since $\varphi_y$ is absolutely continuous, it follows from lemma \ref{lemma:final_delta_fun_bound} that
\begin{align}\label{eq:interim_inequality_delta}
\bigg|\int_a^b \bigg(\pi \delta(x_1 - y_1) - \frac{\sin p(x_1 - y_1)}{x_1 - y_1}\bigg) \varphi_{y_2}(x_1) dx_1\bigg| \leq \xi(a, b, y_1, p) \norm{\varphi_{y_2}}_{[a, b]}^\mathcal{S}.
\end{align}
Furthermore, it also follows from lemma \ref{lemma:final_delta_fun_bound} and the fact that $K$ was absolutely continuous in either of its arguments that
\[
\sup_{x \in [a, b]} |\varphi_{y_2}(x)| \leq \xi(a, b, y_2, p) \norm{K}_{[a, b]^2}^\mathcal{S}.
\]
Similarly, for $x$ almost everywhere in $[a, b]$, $\varphi_{y_2}'(x)$ exists and consequently from lemma \ref{lemma:final_delta_fun_bound} it follows that
\[
\text{For } x \in_{a.e.} [a, b], \ (b - a) |\varphi'_{y_2}(x)| \leq \xi(a, b, y_2, p)\norm{K}_{[a, b]^2}^\mathcal{S}.
\]
Consequently, $\norm{\varphi_{y_2}}^\mathcal{S}_{[a, b]} \leq 2\xi(a, b, y_2, p) \norm{K}_{[a, b]^2}^\mathcal{S}$. Using this along with Eq.~\ref{eq:interim_inequality_delta} completes the proof $\square$.

\begin{proposition}\label{prop:point_coupling_ass1} For $V_1, V_2 \dots V_N \in \mathbb{C}$, $\tau_1, \tau_2 \dots \tau_N \in \mathbb{R}$, define a coupling function $v \in \textnormal{C}_b^\infty(\mathbb{R})$ by $v(\omega) = \sum_{i=1}^N V_i e^{i\omega \tau_i} \ \forall \omega \in \mathbb{R}$, then assumption 1 is satisfied for $v$.
\end{proposition}
\emph{Proof}: We note that it is possible to express $|v|^2$ as a finite sum of harmonics:
\[
|v(\omega)|^2 = \sum_{i=1}^{\tilde{N}} \tilde{V}_i e^{i\omega T_i}.
\]
For any $K \in \text{AC}_\text{sym}([0, t]^2)$, we obtain
\begin{align*}
&\lim_{\Omega \to \infty} \int_{\omega_c\leq |\omega_1|, |\omega_2|\leq \Omega} |v(\omega_1)|^2 |v(\omega_2)|^2 \bigg(\int_{s_1, s_2 = 0}^t K(s_1, s_2) e^{i\omega_2(t - s_2)}e^{-i\omega_1(t - s_1)} ds \bigg)d\omega_1 d\omega_2  \nonumber\\
&=4\bigg|\sum_{i, j = 1}^{\tilde{N}}\tilde{V}_i \tilde{V}_j^* \int_{s_1, s_2 = 0}^t \bigg(\pi\delta(s_2 - t - T_i) - \frac{\sin \omega_c(s_2 - t - T_i)}{s_2 - t - T_i}\bigg)\bigg(\pi \delta(s_1 - t - T_j) - \frac{\sin \omega_c(s_1 - t - T_j)}{s - t - T_j}\bigg)K(s_1, s_2) ds_1 ds_2 \bigg| \nonumber\\
&\leq 4\sum_{i, j = 1}^{\tilde{N}}|\tilde{V}_i \tilde{V}_j^*| \bigg|\int_{s_1, s_2 = 0}^t \bigg(\pi\delta(s_2 - t - T_i) - \frac{\sin \omega_c(s_2 - t - T_i)}{s_2 - t - T_i}\bigg)\bigg(\pi \delta(s_1 - t - T_j) - \frac{\sin \omega_c(s_1 - t - T_j)}{s - t - T_j}\bigg)K(s_1, s_2) ds_1 ds_2 \bigg|.
\end{align*}
Applying lemma \ref{lemma:delta_appx_kernel}, we immediately obtain
\begin{align*}
&\lim_{\Omega \to \infty} \int_{\omega_c\leq |\omega_1|, |\omega_2|\leq \Omega} |v(\omega_1)|^2 |v(\omega_2)|^2 \bigg(\int_{s_1, s_2 = 0}^t K(s_1, s_2) e^{i\omega_2(t - s_2)}e^{-i\omega_1(t - s_1)} ds \bigg)d\omega_1 d\omega_2  \nonumber\\
&\qquad \leq \bigg(\sum_{i, j = 1}^{\tilde{N}} 8\big|\tilde{V}_i \tilde{V}_j\big| \xi(0, t, t + T_i, \omega_c) \xi(0, t, t + T_j, \omega_c)\bigg)\norm{K}_{[0, t]^2}^\mathcal{S},
\end{align*}
which proves that assumption 1 is satisfied $\square$.

We point out that proposition \ref{prop:point_coupling_ass1} shows that assumption 1 is satisfied for both the Markovian model as well as models that model retardation effects.

\subsection{Assumption 2}\label{app:assump2}
In this appendix, we consider two classes of open quantum systems wherein assumption 2 can be proven rigorously. The first is the case of a Markovian coupling function, and the second is the a non-Markovian coupling constant with a bounded square-integrable coupling function. \\ \ \\
\textbf{Assumption 2 (Restated)}
$\forall t\geq 0, s \in [0, t]^{N - 1}$, the map $G_N(\{\cdot, s\}; t): [0, t] \to \mathfrak{L}(\mathcal{H}_S)$ is absolutely continuous and $\exists \gamma(t) > 0$ such that
\begin{align*}
\underset{\{s_0, s\} \in [0, t]^N}{\textnormal{esssup}}\norm{\partial_{s_0} G_N(\{s_0, s\}; t)} \leq  \gamma(t) \norm{L}^N.
\end{align*}

\begin{proposition}\label{prop:assump2_markovian} Assumption 2 is satisfied for a constant coupling function $v$  i.e.~$\forall \omega \in \mathbb{R}: v(\omega) = v_0$ for some constant $v_0$.
\end{proposition}
\emph{Proof}: We begin by noting that constant coupling functions imply Markovian dynamics for the local system, in which case the quantum regression theorem implies that the $N-$point Green's function can be expressed entirely evolving the local system under a non-Hermitian effective Hamiltonian \cite{trivedi2018few, caneva2015quantum}. More specifically, for all $s \in [0, t]^N$
\[
G_N(s; t) = U_\text{eff}(t, 0) \mathcal{T}\big(L_\text{eff}(s_1) L_\text{eff}(s_2) \dots L_\text{eff}(s_N)\big),
\]
where $U_\text{eff}(\tau_1, \tau_2)$ is the propagator corresponding to the Hamiltonian $H_\text{eff}(t) = H_S(t) - i\pi |v_0|^2 L^\dagger L / 2$ and $L_\text{eff}(s) = U_\text{eff}(0, s) L U_\text{eff}(s, 0)$. We note that $\partial_s L_\text{eff}(s)  = -i [L_\text{eff}(s), H_\text{eff}(s)]$ and consequently
\[
\forall s \in [0, t]: \ \norm{\partial_s L_\text{eff}(s)} \leq \sup_{s\in[0, t]}\norm{[L_\text{eff}(s), H_\text{eff}(s)]}.
\]
Noting that for $s \in [0, t]^{N - 1}$, as a function of $s_0\in[0, t]$ $G_N(\{s_0, s\}; t)$ is differentiable at all points except at $s_0 \in s$, we obtain
\[
\norm{\partial_{s_0} G_N(\{s_0, s\}; t)}  \leq \norm{L} ^{N } \gamma(t) \ \text{for } s_0 \in_{a.e.}[0, t],
\]
where
\[
\gamma(t) = \frac{1}{\norm{L}}\sup_{s \in [0, t]} \norm{[L_\text{eff}(s), H_\text{eff}(s)]},
\]
and we have used that since $\norm{U_\text{eff}(\tau_1, \tau_2)}  \leq 1$, $\norm{L_\text{eff}(s)}  \leq \norm{L}  \ \forall \ s\in[0, t]$ $\square$.
\begin{proposition}\label{prop:assump2_sq_int} Consider a coupling function $v \in \textnormal{C}_b^\infty(\mathbb{R})\cap \textnormal{L}^2(\mathbb{R})$, then assumption 2 is satisfied.
\end{proposition}
\noindent\emph{Proof}: Let $s \in [0, t]^{N - 1}$ and without loss of generality, we assume that, $s_1 \leq s_2 \dots  \leq s_{N - 1}$ since the $N-$point Green's function is a symmetric function of its arguments. We consider the $N-$point Green's function $G_N(\{s_0, s\}; t)$ as a function of $s_0$. Consider the case where for some $i \in \{0, 1 \dots N - 2\}, s \in (s_i, s_{i + 1})$ (where $s_0 \equiv -\infty$ and $s_{N} \equiv \infty$) in which case the explicitly applying the time-ordering operator we obtain
\[
G_N(\{s_0, s\}; t) = \bra{\text{vac}}U(t, 0) L(s_{N - 1}) \dots L(s_{i + 1}) L(s_0) L(s_i) \dots L(s_1) \ket{\text{vac}}
\]
and therefore $\forall \ket{\sigma} \in \mathcal{H}_S$
\[
\norm{\partial_{s_0} G_N(\{s_0, s\}; t)\ket{\sigma}}  \leq \norm{L} ^{N - i - 1} \norm{[L, H(s_0)]U(s_0, 0) L(s_i) \dots L(s_1)\ket{\sigma}\otimes\ket{\text{vac}}}.
\]
Noting that 
\[
[L, H(s_0)] = [L, H_S(s_0)] +  [L, L^\dagger] \int_{-\infty}^\infty v(\omega) a_\omega d\omega,
\]
we obtain
\begin{align*}
&\norm{\partial_{s_0} G_N(\{s_0,s\}; t)\ket{\sigma}}  \leq\nonumber\\
&\qquad\norm{L} ^{N - 1} \norm{[L, H_S(s)]}  +\norm{[L, L^\dagger]}  \norm{L} ^{N - i - 1}\norm{\bigg(\int_{-\infty}^\infty v(\omega) a_\omega d\omega \bigg)U(s_0, 0)L(s_i) \dots L(s_1)\ket{\sigma}\otimes\ket{\text{vac}}} .
\end{align*}
To proceed further, we provide a bound on the second term in the above estimate. Denote by $\ket{\phi} = U(s, 0) L(s_i) L(s_{i - 1}) \dots L(s_1) \ket{\sigma}\otimes \ket{\text{vac}}$, we can write $\ket{\phi}$ as a superposition of its $N-$particle components:
\[
\ket{\phi} =\sum_{M = 0}^\infty \ket{\phi^M} = \sum_{M = 0}^\infty \sum_{i=1}^D  \frac{1}{M!} \int_{\omega \in \mathbb{R}^M}  \phi^M_i(\omega) \bigg(\prod_{j = 1}^M a_{\omega_j}^\dagger\bigg) \ket{e_i}\otimes \ket{\text{vac}},
\]
where $\{\ket{e_i}, i \in \{1, 2 \dots D\}\}$ is a basis for the $D-$dimensional local system and
\[
\phi^M_i(\omega) = \bra{e_i}\otimes\bra{\text{vac}}\prod_{i=1}^M a_{\omega_i} \ket{\phi} = \bra{e_i} \otimes \bra{\text{vac}}U(s_0, 0) \mathcal{T}\bigg[\prod_{j=1}^M a_{\omega_j}(s_0)\prod_{j=2}^i L(s_j)\bigg]\ket{\sigma}\otimes \ket{\text{vac}},
\]
where $a_\omega(s) = U(0, s) a_\omega U(s, 0)$ and we have used $s_0 \geq s_{i}\geq s_{i-1} \dots \geq s_1$ to introduce the time-ordering operator. We then use the Heisenberg equations of motion for $a_\omega(s)$ to express
\[
a_\omega(s_0) = a_\omega(0) e^{-i\omega s} - iv^*(\omega) \int_0^{s_0} L(\tau)e^{-i\omega(s_0 - \tau)}d\tau.
\]
We thus obtain
\begin{align*}
\phi_i^M(\omega) &= (-i)^M \bigg(\prod_{j =1}^M v^*(\omega_j)\bigg) \int_{\tau \in [0, s]^M}\bra{e_i} \otimes \bra{\text{vac}}U(s, 0) \mathcal{T}\bigg[\prod_{j=1}^M L(\tau_j)\prod_{j=2}^i L(s_j)\bigg]\ket{\sigma}\otimes \ket{\text{vac}} e^{-i\omega\cdot(s - \tau)}d\tau \nonumber\\
&=(-i)^M \bigg(\prod_{j=1}^M v^*(\omega_j)\bigg) \int_{\tau\in[0, s]^M} \bra{e_i}G_{M + i}([\tau, s]; s_0)\ket{\sigma}e^{-i\omega \cdot (s_0 - \tau)}d\tau.
\end{align*}
This immediately allows us to bound the norm of the $M-$particle component of $\ket{\phi}$:
\[
\norm{\ket{\phi^M}}^2 \leq  \frac{(2\pi)^M\norm{v}_\infty^{2M}}{M!}\int_{\tau \in [0, s]^M}\bra{\sigma}G_{M + i}^\dagger([\tau, s_2 \dots s_i]; s)G_{M + i}([\tau, s_2 \dots s_i]; s) \ket{\sigma} d\tau \leq \frac{\norm{v}_\infty^{2M}(2\pi s)^M}{M!} \norm{L} ^{2(M + i)}.
\]
Finally, noting that the restriction of the operator $\int_{-\infty}^\infty v(\omega) a_\omega d\omega$ on the $M-$particle subspace is $\sqrt{M} \big[\int_{-\infty}^\infty |v(\omega)|^2d\omega\big]^{1/2}$, we obtain that
\begin{align*}
\norm{\bigg(\int_{-\infty}^\infty v(\omega) a_\omega d\omega\bigg)U(s, 0) L(s_{i - 1}) \dots L(s_2) \ket{\sigma}\otimes \ket{\text{vac}} }^2 &\leq \bigg(\int_{-\infty}^\infty |v(\omega)|^2d\omega\bigg) \sum_{M = 0}^\infty  M \norm{\ket{\phi^M}}^2 \nonumber\\
&\leq \norm{L} ^{2(i + 1)} 2\pi s\norm{v}_\infty^2 e^{2\pi\norm{v}_\infty^2 \norm{L} ^2s },
\end{align*}
and consequently we obtain that $\forall \sigma \in \mathcal{H}_S, s_0 \in (s_i, s_{i + 1})$,
\[
\norm{\partial_{s_0}G_N(\{s_0, s\}; t) \ket{\sigma}}  \leq  \norm{L} ^N \bigg(\sup_{s_0 \in [0, t]}\frac{\norm{[L, H_S(s_0)]} }{\norm{L} } + \norm{[L, L^\dagger]} \sqrt{2\pi t} \norm{v}_\infty e^{\pi \norm{v}_\infty^2 \norm{L} ^2 t/ 2} \int_{-\infty}^\infty |v(\omega)|^2d\omega \bigg).
\]
Since this bound holds for every $i \in \{0, 1 \dots N - 2\}$, it proves that assumption 2 holds for square integrable coupling functions $\square$.

\section{Detailed Proofs}\label{sec:detailed_proofs}
\subsection{Proof of theorem 1: Introducing a frequency cut-off on the environment}
We first show rigorously that, subject to the assumptions 1 and 2, a frequency cutoff can be introduced in the environment without significantly impacting the dynamics of the local system.

\edit{\begin{definition}
\label{def:projector}
For a given frequency cutoff $\omega_c >0$, let $\Pi_{\omega_c}:  \mathcal{H}_E \to  \mathcal{H}_E$ such that $\forall N \geq 0, \psi \in \textnormal{L}^2(\mathbb{R}^N)$
\begin{align}
 \Pi_{\omega_c} \int_{\omega \in \mathbb{R}^N} \psi(\omega) \bigg[\prod_{i=1}^N a_{\omega_i}^\dagger\bigg]\ket{\textnormal{vac}} d\omega = \int_{\omega \in [-\omega_c, \omega_c]^N} \psi(\omega) \bigg[\prod_{i=1}^N a_{\omega_i}^\dagger\bigg] \ket{\textnormal{vac}} d\omega.
\end{align}
Furthermore, $\mathcal{P}_{\omega_c}, \mathcal{Q}_{\omega_c}:\mathcal{H}_S\otimes \mathcal{H}_E \to \mathcal{H}_S\otimes \mathcal{H}_E$ such that $\mathcal{P}_{\omega_c} = \textnormal{id}\otimes \Pi_{\omega_c}$ and $\mathcal{Q}_{\omega_c} = \textnormal{id} - \mathcal{P}_{\omega_c}$.
\end{definition}}
The operator $\mathcal{P}_{\omega_c}$ thus projects a given system-environment state onto the space of states with no environment frequencies outside of $|\omega| \leq \omega_c$. It is straightforward to see that the Hamiltonian $H_{\omega_c}(t)$ defined in theorem 1 satisfy $H_{\omega_c}(t) = \mathcal{P}_{\omega_c} H(t) \mathcal{P}_{\omega_c}$. We now study the error introduced by evolving an initial state under the Hamiltonian $H_{\omega_c}(t)$ instead of the Hamiltonian $H(t)$. This is made explicit in the following lemma.
\edit{\begin{lemma} \label{lemma:decomp}Given an initial state $\ket{\psi_0}$, and denoting by $U(t_1, t_2)$ and $U_{\omega_c}(t_1, t_2)$ the propagators corresponding to $H(t)$ and $H_{\omega_c}(t)$ respectively, the states $\ket{\psi(t)} = U(t, 0)\ket{\psi_0}$ and $\ket{\psi_{\omega_c}(t)} =U_{\omega_c}(t, 0)\mathcal{P}_{\omega_c}\ket{\psi_0}$ satisfy
\begin{align*}
\norm{\ket{\psi(t)} - \ket{\psi_{\omega_c}(t)}} \leq \norm{\mathcal{Q}_{\omega_c} \ket{\psi_0}} + \norm{\mathcal{Q}_{\omega_c}\ket{\psi(t)}} + \int_0^t \norm{\mathcal{P}_{\omega_c} H(s) \mathcal{Q}_{\omega_c} \ket{\psi(s)}}ds
\end{align*}
\end{lemma}
\emph{Proof}: Note that $\ket{\psi(t)} = \mathcal{P}_{\omega_c}\ket{\psi(t)} + \mathcal{Q}_{\omega_c}\ket{\psi(t)}$. From the Schroedinger's equation for $\ket{\psi(t)}$, it then follows that
\begin{align*}
i\frac{d}{dt}\mathcal{P}_{\omega_c}\ket{\psi(t)} = \mathcal{P}_{\omega_c}H(t) \ket{\psi(t)} =H_{\omega_c}(t) \mathcal{P}_{\omega_c}\ket{\psi(t)} + \mathcal{P}_{\omega_c}H(t) \mathcal{Q}_{\omega_c} \ket{\psi(t)},
\end{align*}
which can be integrated to obtain
\begin{align*}
\mathcal{P}_{\omega_c}\ket{\psi(t)} =U_{\omega_c}(t, 0) \mathcal{Q}_{\omega_c}\ket{\psi_0} + \ket{\psi_{\omega_c}(t)}- i\int_0^t U_{\omega_c}(t, s)\mathcal{P}_{\omega_c}H(s) \mathcal{Q}_{\omega_c} \ket{\psi(s)}ds.
\end{align*}
Consequently
\begin{align*}
\ket{\psi(t)} - \ket{\psi_{\omega_c}(t)} =U_{\omega_c}(t, 0)\mathcal{Q}_{\omega_c}\ket{\psi_0} +  \mathcal{Q}_{\omega_c} \ket{\psi(t)} - i\int_0^t U_{\omega_c}(t, s)\mathcal{P}_{\omega_c}H(s) \mathcal{Q}_{\omega_c} \ket{\psi(s)}ds,
\end{align*}
from which the statement of the lemma follows. $\square$}

The two terms in the error estimate provided in this lemma can be respectively interpreted as the error introduced in quantum state of the system on ignoring the high-frequency components of the environment state, and the error introduced by not accounting for the coupling of the high-frequency components of the environment state to the low-frequency components by virtue of their interaction with the local system. In the following lemmas, we show that both of these errors are bounded by a function that goes to 0 as the frequency cutoff $\omega_c$ is made increasingly larger.
\begin{lemma}\label{lemma:error_direct} If assumptions 1 and 2 are satisfied, it follows that $\forall t > 0$,
\begin{align*}
\norm{ \mathcal{Q}_{\omega_c}\ket{\psi(t)}}  \leq \frac{\norm{v}_\infty\norm{L} }{\sqrt{\omega_c}} (2 + \gamma(t) t)  e^{\pi\norm{v}_\infty^2\norm{L} ^2 t },
\end{align*}
where $\gamma(t)$ is introduced in assumption 2.
\end{lemma}
\noindent\emph{Proof}: It follows from the definition of $\mathcal{Q}_{\omega_c}$ that
\[
\norm{\mathcal{Q}_{\omega_c}\ket{\psi(t)}}  = \sqrt{\bra{\psi(t)}(\textnormal{id}-\mathcal{P}_{\omega_c})\ket{\psi(t)}}.
\]
Using Eq.~\ref{eq:N_photon_component}, we obtain
\begin{align*}
\bra{\psi(t)}\psi(t)\rangle = \sum_{N=0}^\infty \frac{1}{N!}\int_{ \mathbb{R}^N}\bra{\sigma}F_N^\dagger(\omega; t) F_N(\omega; t)\ket{\sigma}\prod_{i=1}^N |v(\omega_i)|^2 d \omega
\end{align*}
and
\begin{align}
&\bra{\psi(t)}\mathcal{P}_{\omega_c}\ket{\psi(t)} = \sum_{N = 0}^{\infty} \frac{1}{N!} \int_{[-\omega_c, \omega_c]^N} \bra{\sigma}F_N^\dagger(\omega; t) F_N(\omega; t) \ket{\sigma} \prod_{i=1}^N |v(\omega_i)|^2d\omega.
\end{align}
Therefore,
\begin{align*}
&\bra{\psi(t)}\mathcal{Q}_{\omega_c}\ket{\psi(t)}   \leq \sum_{N=1}^{\infty}\frac{\norm{v}_\infty^{2N}}{(N - 1)!} \int_{\substack{|\omega_1|\geq \omega_c \\ \omega_2\dots \omega_N \in \mathbb{R}}}  \bra{\sigma}F_N^\dagger(\omega; t) F_N(\omega; t)\ket{\sigma} d \omega \nonumber\\ 
&\qquad \qquad \qquad \quad=\sum_{N=1}^{\infty}\frac{(2\pi \norm{v}_\infty^2)^{N}}{2\pi(N - 1)!} \int_{\substack{|\omega|\geq \omega_c \\ s\in [0, t]^{N-1}}}  \bra{\sigma}H_N^\dagger(\omega, s; t) H_N(\omega, s; t)\ket{\sigma}d\omega\ ds.
\end{align*} 
where, for $N\geq 1$ and $s\in [0, t]^{N-1}$, $H_N(\omega, s; t)$ is given by
\begin{align*}
H_N(\omega, s; t) = \int_{s' \in [0, t]} G_N([s', s]; t) e^{-i\omega (t - s')} ds'.
\end{align*}
Using integration by parts, together with assumption 2, we immediately obtain that
\begin{align*}
\norm{H_N(\omega, s;t)}  \leq \frac{(2 + \gamma(t)t)}{\omega} \norm{L} ^N,
\end{align*}
and therefore
\begin{align*}
\bra{\psi(t)}\mathcal{Q}_{\omega_c}\ket{\psi(t)} \leq \frac{\norm{v}_\infty^2 \norm{L} ^2}{\omega_c} (2 + \gamma(t) t)^2 e^{2\pi \norm{v}_\infty^2 \norm{L} ^2 t }.
\end{align*}
from which the lemma immediately follows $\square$.

\begin{lemma}\label{lemma:error_coup} If assumptions 1 and 2 are satisfied, it follows that $\forall t > 0$,
\begin{align*}
&\int_0^t \norm{\mathcal{P}_{\omega_c} H(\tau) \mathcal{Q}_{\omega_c}\ket{\psi(\tau)}}  d\tau \leq \norm{L} ^2 \int_0^t \tau (1 + \gamma(\tau) \tau) \big[V(\omega_c, \tau)\big]^{1/2}e^{\pi\norm{v}_\infty^2 \norm{L} ^2 \tau }d\tau,
\end{align*}
where $\gamma(t)$ is introduced in assumption 2.
\end{lemma}
\emph{Proof}: Using the definitions of $\mathcal{P}_{\omega_c}$ and $\mathcal{Q}_{\omega_c}$, it follows that
\begin{align*}
\mathcal{P}_{\omega_c} H(\tau) \mathcal{Q}_{\omega_c}\ket{\psi(\tau)} =  L^\dagger \sum_{N=1}^{\infty}\frac{1}{(N - 1)!}\int_{\substack{|\omega_1|\geq \omega_c \\ \omega_2 \dots \omega_N \in \mathbb{R}}} |v(\omega_1)|^2 F_N(\omega; \tau)\bigg[\prod_{i=2}^{N} v^*(\omega_i)a_{\omega_i}^\dagger\bigg]\ket{\sigma}\otimes \ket{\text{vac}} d\omega.
\end{align*}
Therefore,
\begin{align*}
&\norm{\mathcal{P}_{\omega_c} H(\tau) \mathcal{Q}_{\omega_c}\ket{\psi(\tau)}}^2 \leq \sum_{N=1}^{\infty} \frac{(2\pi \norm{v}_\infty^2)^{N - 1}}{(N - 1)!} \int_{\substack{s \in [0, t]^{N-1} \\ |\omega|, |\omega'|\geq \omega_c}}  |v(\omega)|^2 |v(\omega')|^2 \bra{\sigma}H_N^\dagger(\omega', s; \tau) LL^\dagger H_N(\omega, s; \tau)\ket{\sigma} d\omega\ d\omega' ds. \nonumber 
\end{align*}
To proceed further, we make use of assumption 1. For $\tau > 0, N\geq 1, \ s \in [0, t]^{N-1}$, we construct a positive kernel $K\in \text{AC}_\text{sym}([0, \tau]^2)$ from the $N-$point Green's function via $K(t', t'')= \bra{\sigma}G_N^\dagger(\{t', s\}; \tau) LL^\dagger G_N(\{t'', s\}; \tau)\ket{\sigma} $. An application of the assumption 1 using this kernel yields $\forall s \in [0, t]^{N-1}$\
\begin{align*}
&\int_{|\omega|, |\omega'| \geq\omega_c} |v(\omega)|^2 |v(\omega')|^2 \bra{\sigma}H_N^\dagger(\omega', s;\tau) LL^\dagger H_N(\omega, s;\tau)\ket{\sigma} d\omega' d\omega \leq V(\omega_c, \tau) \norm{\bra{\sigma}G_N^\dagger(\{\cdot, s\}; \tau) LL^\dagger G_N(\{\cdot, s\}; \tau)\ket{\sigma}}_{[0, \tau]^2}^{\mathcal{S}},
\end{align*}
where $V(\omega_c, t)$ is the function introduced in assumption 1. Using assumption 2, we immediately obtain that $\forall s \in [0, \tau]^{N-1}$
\begin{align*}
\norm{\bra{\sigma}G_N^\dagger([\cdot, s]; \tau) LL^\dagger G_N([\cdot, s]; \tau)\ket{\sigma}}_{[0, t]^2}^{\mathcal{S}} \leq  \norm{L} ^{2N + 2} (1 + \gamma(\tau) \tau)^2
\end{align*}
It thus follows that
\begin{align*}
\norm{\mathcal{P}_{\omega_c}H(\tau)\mathcal{Q}_{\omega_c}\ket{\psi(\tau)}}\leq \norm{L} ^2 (1 + \gamma(\tau) \tau) e^{\pi\norm{v}_\infty^2\norm{L} ^2\tau}
\end{align*}
Using this bound, the lemma statement follows. $\square$ \\

\noindent\textbf{Theorem 1 (Restated)} \emph{Suppose $v \in \textnormal{C}_b^\infty(\mathbb{R})$ is a coupling function such that assumptions 1 and 2 are satisfied. Denoting by $\rho(t)$ the reduced density matrix of the local system at time $t$ when an initial state $\ket{\sigma}\otimes\ket{\text{vac}}$ is evolved under the Hamiltonian in Eq.~1 and by ${\rho}_{\omega_c}(t)$ the reduced density matrix of the local system at time $t$ when the same initial state is evolved under the Hamiltonian
\begin{align}\label{eq:hamiltonian_with_cutoff}
H_{\omega_c}(t) =& H_S(t) + \int_{-\infty}^{\infty} \omega a^\dagger_\omega a_\omega d\omega + \int_{-\omega_c}^{\omega_c} \bigg( v(\omega) a_\omega L^\dagger + v^*(\omega) a^\dagger_\omega L\bigg)d\omega,
\end{align}
then
\begin{align*}
\norm{\rho(t) - {\rho}_{\omega_c}(t)}_\textnormal{tr} \leq \varepsilon(\omega_c, t),
\end{align*}
where $\varepsilon(\omega_c, t)$ is the cutoff error given by
\begin{align}
\varepsilon(\omega_c, t) = \frac{f_1(t)}{\sqrt{\omega_c}} + \int_0^t f_2(\tau) \sqrt{V(\omega_c, \tau)}d\tau.
\end{align}
Here $V(\omega_c, t)$ is defined in assumption 1 and
\begin{align*}
&f_1(\tau)= \sqrt{2}\norm{v}_\infty\norm{L} ( 2+ \gamma(\tau) \tau) e^{\pi\norm{v}_\infty^2 \norm{L}^2\tau}, f_2(\tau) = \sqrt{2}\norm{L}^2 \tau (1 + \gamma(\tau) \tau) e^{\pi\norm{v}_\infty^2 \norm{L}^2 \tau },
\end{align*}
where $\gamma(t)$ is introduced in assumption 2.}

\noindent \emph{Proof}: Let $\ket{\sigma} \in \mathcal{H}_S$ be an arbitrary local system state, and denote by $\ket{\psi(t)}$ the joint state of the system and the environment obtained on evolving $\ket{ \sigma} \otimes \ket{\text{vac}}$ under the Hamiltonian $H(t)$. Similarly, we denote by $\ket{\psi_{\omega_c}(t)}$ the joint state of the system and environment obtained on evolving $\ket{\sigma}\otimes\ket{\text{vac}}$ under the Hamiltonian $H_{\omega_c}(t) = \mathcal{P}_{\omega_c} H(t) \mathcal{P}_{\omega_c}$. We denote by $\rho(t)$ and $\rho_{\omega_c}(t)$ the reduced state of the local system at time $t$ i.e. $\rho(t) = \text{Tr}_{\mathcal{H}_B}[\ket{\psi(t)}\bra{\psi(t)}]$ and $\rho_{\omega_c}(t) = \text{Tr}_{\mathcal{H}_B}[\ket{\psi(t)}\bra{\psi(t)}]$. We can then use the contractivity of the partial trace, as well as the relationship between the trace-norm distance and the norm distance between two pure states (Eq.~\ref{eq:tr_norm_l2}), we obtain
\begin{align*}
\norm{\rho(t) - \rho_{\omega_c}(t)}_\text{tr}&\leq \norm{\ket{\psi(t)}\bra{\psi(t)} - \ket{\psi_{\omega_c}(t)}\bra{\psi_{\omega_c}(t)}}_\text{tr} \nonumber \\
&\leq \sqrt{2}\norm{\ket{\psi(t)} - \ket{\psi_{\omega_c}(t)}}.
\end{align*}
An application of lemmas \ref{lemma:decomp}, \ref{lemma:error_direct} and \ref{lemma:error_coup} allows us to bound $\norm{\ket{\psi(t)} - \ket{\psi_{\omega_c}(t)}}$ and obtain the result in theorem $\square$.

\subsection{Proof of theorem 2: Convergence of pseudomode approximation}
\noindent\textbf{Definition 3 (Pseudo-mode description, restated)}\emph{ An environment described by $M-$pseudomodes with parameters $\{(\omega_i, g_i, \kappa_i \geq 0) : i \in \{0, 1, 2 \dots M - 1\}\}$ has an associated Hilbert space $\mathcal{H}_\textnormal{aux}$ of $M$ bosonic modes with annihilation operators $a_0, a_1 \dots a_{M - 1}$. For a local system with Hilbert space $\mathcal{H}_S = \mathbb{C}^d$, time-dependent Hamiltonian $H_S(t) \in \mathfrak{L}(\mathbb{C}^d)$, interacting with the environment through the operator $L \in \mathfrak{L}(\mathbb{C}^d)$ with the initial system-environment state being $\ket{\sigma}\otimes \ket{\textnormal{vac}}$, its reduced state at time $t$ is given by $\rho(t) = \textnormal{Tr}_\textnormal{aux}(R(t))$, where $R(t)$ satisfies
\begin{align*}
\dot{R}(t) = i[\hat{H}(t), R(t)] + \sum_{i=0}^{M - 1} \frac{\kappa_i}{2}\big(2a_i R(t) a_i^\dagger - \{a_i^\dagger a_i, R(t)\}\big),
\end{align*}
with
\begin{align*}
\hat{H}(t) =  H_S(t) + \sum_{i=0}^{M - 1} \omega_i a_i^\dagger a_i + \sum_{i=1}^M g_i(a_i L^\dagger + a_i^\dagger L),
\end{align*}
and $R(0) = \ket{\sigma}\bra{\sigma} \otimes (\ket{0}\bra{0})^{\otimes M}$.}

\begin{lemma}[Ref.~\cite{tamascelli2018nonperturbative}]
\label{lemma:pseudo_mode_appx}
An environment interacting with a local system with a coupling function $v$ which satisfies 
\begin{align}
|v(\omega)|^2 = \sum_{i=1}^M\frac{\kappa_i}{2\pi} \frac{g_i^2}{(\omega - \omega_i)^2 + \kappa_i^2 / 4} \ \forall \ \omega \in \mathbb{R},
\end{align}
for some $\{(\omega_i, g_i, \kappa_i > 0) : i \in \{0, 1, 2 \dots M - 1\}\}$, then the local system dynamics is exactly described by a pseudomode description (definition 3) with parameters $\{(\omega_i, g_i, \kappa_i) : i \in \{0, 1, 2 \dots M - 1\}\}$.
\end{lemma}
We first develop bounds on the error incurred on approximating one bounded and square integrable coupling constant with another one which are precisely stated in the following lemma.
\begin{lemma} \label{lemma:error_state}Let $v_1, v_2 \in \textnormal{C}^\infty_b(\mathbb{R}) \cap \textnormal{L}^2(\mathbb{R})$ be coupling functions and let $H_1(t), H_2(t)$ denote the Hamiltonians corresponding to these coupling functions as given by Eq.~1 in the main text.
Denoting by $\ket{\psi_i(t)}$ as the joint state of the local system and environment when an initial state $\ket{\sigma}\otimes \ket{\textnormal{vac}}$, with $\ket{\sigma} \in \mathcal{H}_S$, is evolved under the Hamiltonian $H_i(t), \ i \in \{1, 2\}$, then
\begin{align}
\norm{\ket{\psi_1(t)} - \ket{\psi_2(t)}} \leq g(t)d_{\omega_c}(v_1, v_2),
\end{align}
where
\begin{align*}
g(t) = \int_0^t \bigg(\sqrt{1 + \norm{v_1}_\infty^2 \tau \norm{L} ^2 e^{2\pi\norm{v_1}_\infty^2 \norm{L} ^2 \tau}} + \norm{v_1}_\infty \sqrt{\tau} \norm{L}  e^{\pi\norm{v_1}^2_\infty \norm{L} ^2 \tau }\bigg)d\tau
\end{align*}
and
\[
d(v_1, v_2) = \bigg[\int_{-\infty}^\infty |v_1(\omega) - v_2(\omega)|^2d\omega\bigg]^{1/2},
\]
quantifies a distance between the two coupling functions.
\end{lemma}
\noindent \emph{Proof}: For $i \in \{1, 2\}$, we denote by $U_i(\tau_1, \tau_2)$ the propagator corresponding to the Hamiltonian $H_i$. Furthermore, let $\ket{\psi_i(t)} = U_i(t, 0)\ket{\sigma}\otimes \ket{\text{vac}}$, where $\ket{\sigma} \in \mathcal{H}_S$. It can be seen that
\[
\norm{\ket{\psi_1(t)} - \ket{\psi_2(t)}} = \norm{\ket{\phi(t)} - \ket{\sigma}\otimes \ket{\text{vac}}},
\]
where
\[
\ket{\phi(t)} = U_2^\dagger(t, 0)\ket{\psi_1(t)}.
\]
An integral equation can easily be derived by differentiating this equation, followed by integrating it from $0$ to $t$ to obtain
\[
\ket{\phi(t)} = \ket{\sigma}\otimes \ket{\text{vac}} + i \int_0^tU_2(t, \tau)(H_2(\tau) - H_1(\tau)) \ket{\psi_1(\tau)}d\tau,
\]
from which we obtain the bound
\[
\norm{\ket{\psi_1(t)} - \ket{\psi_2(t)}} \leq \int_0^t \norm{(H_2(\tau) - H_1(\tau))\ket{\psi_1(\tau)}}d\tau.
\]
We note from the definition of the Hamiltonians $H_i(t)$ that $H_2(t) - H_1(t) = \Delta^+L + L^\dagger\Delta^-$ where
\begin{align*}
\Delta^+ = \int_{-\infty}^{\infty} \big(v_1^*(\omega) - v_2^*(\omega)\big) a_\omega^\dagger  d\omega, \nonumber\\ \Delta^- = \int_{-\infty}^{\infty} \big(v_1(\omega) - v_2(\omega)\big) a_\omega d\omega.
\end{align*}
Therefore,
\begin{align}\label{eq:first_estimate}
&\norm{\ket{\psi_1(t)} - \ket{\psi_2(t)}} \leq \nonumber\\
 &\qquad\norm{L} \bigg(\int_0^t \norm{\Delta^+\ket{\psi_1(\tau)}}d\tau + \int_0^t \norm{\Delta^-\ket{\psi_1(\tau)}}d\tau \bigg).
\end{align}
Note that the operators $\Delta^\pm$ are unbounded, and consequently the norms in the above equations cannot be estimated trivially by introducing the operator norms of $\Delta^\pm$. However, $\Delta^\pm$ are bounded when restricted to only the $N-$particle subspaces of the environment's Hilbert space. In particular, for a $N-$particle state $\ket{\psi^N}$
\begin{align*}
&\norm{\Delta^+ \ket{\psi^N}}^2 \leq \norm{\ket{\psi^N}}^2 (N + 1)d^2(v_1, v_2), \\
&\norm{\Delta^- \ket{\psi^N}}^2 \leq \norm{\ket{\psi^N}}^2N d^2(v_1, v_2).
\end{align*}
Furthermore, the $N-$particle component of the state $\ket{\psi_1(t)}$ has a bounded norm, with Eq.~5 of the main text providing the explicit bound. It thus immediately follows that
\begin{align*}
&\norm{\Delta^+ \ket{\psi_1(t)}}^2 \leq \big(1 + \norm{v_1}_\infty^2 t \norm{L} ^2 e^{2\pi\norm{v_1}_\infty^2 \norm{L} ^2 t}\big) d^2(v_1, v_2) \\
&\norm{\Delta^- \ket{\psi_1(t)}}^2 \leq \norm{v_1}_\infty^2 t \norm{L} ^2 e^{2\pi\norm{v_1}_\infty^2 \norm{L} ^2 t} d^2(v_1, v_2)
\end{align*} 
These estimates, along with Eq.~\ref{eq:first_estimate}, proves the statement of the lemma. $\square$.

While this estimate allows for an assessment of the error incurred in approximating one coupling function with another, if only the state of the local system is of interest, the phase of the coupling function is irrelevant. This is easily seen from the fact that the transformation $a_\omega \to a_\omega e^{i\theta(\omega)}$ and $v(\omega) \to v(\omega) e^{-i\theta(\omega)}$, for any $\theta: \mathbb{R}\to\mathbb{R}$, leaves both the Hamiltonian in Eq.~1 of the main text and the initial state (provided the environment is in the vacuum state) unchanged. Consequently, a bound can be provided on the error in the reduced density matrix of the local system that only depends on the magnitude of the coupling constant, as made explicit by the following lemma.

\begin{lemma}\label{lemma:final_appx_error}
Let $v_1, v_2 \in \textnormal{C}^\infty_b(\mathbb{R}) \cap \textnormal{L}^2(\mathbb{R})$ be coupling functions and let $H_1(t), H_2(t)$ denote the Hamiltonians corresponding to these coupling functions as given by Eq.~1 of the main text.
Denoting by $\rho_i(t)$ as the reduced state of the local system when an initial state $\ket{\sigma}\otimes \ket{\textnormal{vac}}$, with $\ket{\sigma} \in \mathcal{H}_S$, is evolved under the Hamiltonian $H_i(t), \ i \in \{1, 2\}$, then
\begin{align*}
\norm{\rho_1(t) - \rho_2(t)}_\textnormal{tr} \leq \sqrt{2}g(t) \bigg[\int_{-\omega_c}^{\omega_c} \big| |v_1(\omega)|^2 - |v_2(\omega)|^2\big| d\omega\bigg]^{1/2},
\end{align*}
where $g(t)$ is the function defined in lemma \ref{lemma:error_state}.
\end{lemma}
\emph{Proof}: Denote by $\theta_i(\omega)$ the phase of the coupling coefficient $v_i(\omega) \ \forall \omega \in \mathbb{R}, i\in\{1, 2\}$. Applying the transformation $a_\omega \to a_\omega e^{-i\theta_i(\omega)}$ to the Hamiltonian $H_i$, we obtain
\begin{align*}
\tilde{H}_i(t) = H_S(t) + \int_{-\infty}^{\infty} \omega a_\omega^\dagger a_\omega d\omega + \int_{-\infty}^{\infty}\big|v_i(\omega)\big|\big(a_\omega L^\dagger + a_\omega^\dagger L\big)d\omega.
\end{align*}
For $i \in \{1, 2\}$, denote by $\ket{\psi_i(t)}$ the state obtained by evolving $\ket{\sigma}\otimes \ket{\text{vac}}$ under the Hamiltonian $\tilde{H}_i(t)$. While these states, in their entirety, will be different from the ones obtained on evolving the system under the Hamiltonian $H_i(t), i \in \{1, 2\}$, the reduced density matrix of the local system will be the same. Consequently, we can bound the error in the reduced density matrix of the local system by simply bounding the error between the states $\ket{{\psi_1(t)}}$ and $\ket{{\psi_2(t)}}$, which can be done with an application of lemma \ref{lemma:error_state}:
\begin{align*}
\norm{\ket{\psi_1(t)} - \ket{\psi_2(t)}} \leq g(t) d(|v_1|, |v_2|),
\end{align*}
where $g(t)$ and $d(v_1, v_2)$ are defined in lemma \ref{lemma:error_state}. Furthermore, we note that
\[
d^2(|v_1|, |v_2|) \leq \int_{-\infty}^{\infty}\big| |v_1(\omega)|^2 - |v_2(\omega)|^2\big | d\omega,
\]
and thus we obtain
\[
\norm{\ket{\psi_1(t)} - \ket{\psi_2(t)}} \leq g(t) \bigg[\int_{-\infty}^{\infty} \big| |v_1(\omega)|^2 - |v_2(\omega)|^2\big|d\omega\bigg]^{1/2}
\]
Finally, using Eq.~\ref{eq:tr_norm_l2} together with the fact that partial trace is contractive, we obtain the bound provided in the lemma $\square$.

To use pseudomodes to simulate the dynamics of a non-Markovian quantum system, we would need to approximate the magnitude-square of the coupling function, $|v(\omega)|^2$, by a sum of Lorentzian. Since, as shown in theorem 1, the dynamics of local system is only impacted by the coupling function within a finite frequency window, a natural procedure to obtain the sum of Lorentzian decomposition would be to choose a frequency cutoff $\omega_c > 0$, an integer $M$ indicating the number of Lorentzians and solve the following optimization problem:
\begin{align*}
\min_{\substack{\omega \in \mathbb{R}^M \\  V, \kappa \in \mathbb{R}^{+M}}} \int_{-\infty}^{\infty}\bigg| |v(\omega)|^2 \textnormal{rect}_{\omega_c}(\omega) - \sum_{i = 1}^M \frac{V_i}{(\omega - \omega_i)^2 + \kappa_i^2 / 4}\bigg| d\omega.
\end{align*}
While in practice this optimization problem can be solved numerically, a theoretical analysis of the resulting approximation error as quantified by the solution of this problem is hard since it is a non-convex problem. However, as is made concrete in the following lemma, an upper bound on the approximation error can be provided by an explicit Lorentzian construction.
\begin{lemma}\label{lemma:lemma_appx}Given a differentiable function $\Gamma:\mathbb{R}\to[0, \infty)$ which is also bounded, then for $\omega_c, \gamma \in (0, \infty)$, $M \in \mathbb{N}$,
\begin{align*}
\int_{-\infty}^{\infty} \bigg | \Gamma(\omega) \textnormal{rect}_{\omega_c}(\omega) - \hat{\Gamma}(\omega)\bigg| d\omega \leq \varepsilon(\omega_c, \kappa, M)
\end{align*}
where $\hat{\Gamma}:\mathbb{R}\to\mathbb{R}$ is a lorentzian approximation to $\Gamma$ given by
\[
\hat{\Gamma}(\omega) = \sum_{n=1}^M \frac{\kappa \delta}{\pi} \frac{\Gamma(\omega_n)}{(\omega - \omega_n)^2 + \kappa^2},
\]
with $\delta = 2\omega_c / (M - 1)$ and $\omega_n = -\omega_c + (n - 1)\delta$ and $\varepsilon(\omega_c, \kappa, M)$ is the approximation error which satisfies the scaling
\begin{align*}
&\varepsilon(\omega_c, \kappa, M) = {O}\bigg(\Gamma_\textnormal{max}\kappa\log\frac{\kappa}{\omega_c}\bigg) + {O}\bigg(\omega_c \kappa \Gamma^d_\textnormal{max}(\omega_c)\log\frac{\kappa}{\omega_c}\bigg) + {O}\bigg(\frac{\omega_c^3}{\kappa N}\Gamma^d_\textnormal{max}(\omega_c)\bigg) + {O}\bigg(\frac{\omega_c^3}{\kappa^2 N}\Gamma_\textnormal{max}\bigg) + {O}\bigg(\frac{\Gamma_\textnormal{max}}{M}\bigg),
\end{align*}
where $\Gamma_\textnormal{max} = \sup_{\omega\in\mathbb{R}}\Gamma(\omega)$ and $\Gamma_\textnormal{max}^d(\omega_c) = \sup_{\omega \in [-\omega_c, \omega_c]} |\Gamma'(\omega)|$.
\end{lemma}
\noindent\emph{Proof}: Our goal is to develop a Lorentzian approximation to a function $\Gamma$ which is bounded and differentiable. We develop such an approximation in two steps:
\begin{enumerate}
\item We first use the fact that Lorentzians are mollifiers to construct an approximation $\Gamma_1:\mathbb{R}\to\mathbb{R}$ to $\Gamma$ within the frequency window $[-\omega_c, \omega_c]$ 
\begin{align}\label{eq:mollifier_appx}
\Gamma_1(\omega) = \frac{\kappa}{\pi}\int_{-\omega_c}^{\omega_c}\frac{\Gamma(\nu)}{(\omega - \nu)^2 + \kappa^2}d\nu
\end{align}
Note that as $\kappa \to 0$, $\Gamma_1(\omega)$ is expected to converge to $\Gamma(\omega)$ for $\omega \in (-\omega_c, \omega_c)$. For a non-zero value of $\kappa$, an error will be incurred in approximating $\Gamma$ by $\Gamma_1$.
\item Next, we numerically discretize the integral in Eq.~\ref{eq:mollifier_appx} to obtain the approximation $\hat{\Gamma}:\mathbb{R}\to \mathbb{R}$ to $\Gamma$. Denoting by the number of Lorentzians (i.e.~number of points used in discretizing the integral) by $M$, the discretization used in approximating the integral is $\delta = 2\omega_c / (M - 1)$ and thus
\begin{align}
\hat{\Gamma}(\omega) = \sum_{n = 1}^M \frac{\kappa \delta}{\pi} \frac{\Gamma(\omega_n)}{(\omega - \omega_n)^2 + \kappa^2}.
\end{align}
\end{enumerate}
We now consider the error incurred in approximating $\Gamma$ within a frequency window $[-\omega_c, \omega_c]$ by $\hat{\Gamma}$. The relevant error of interest is the $\text{L}^1$ norm error between $\Gamma\cdot\text{rect}_{\omega_c}$ and $\hat{\Gamma}$, which can be upper bounded by
\[
\int_{-\infty}^\infty \bigg| \Gamma(\omega) \text{rect}_{\omega_c}(\omega) - \hat{\Gamma}(\omega)\bigg|d\omega \leq  \underbrace{\int_{-\omega_c}^{\omega_c} \big|\Gamma(\omega) - \Gamma_1(\omega)\big|d\omega}_\textnormal{Error between $\Gamma$ and $\Gamma_1$} + \underbrace{\int_{-\omega_c}^{\omega_c} \big|\Gamma_1(\omega) - \hat{\Gamma}(\omega)\big|d\omega}_\textnormal{Error between $\Gamma_1$ and $\hat{\Gamma}$} + \underbrace{\int_{|\omega|\geq\omega_c} |\hat{\Gamma}(\omega)|d\omega}_\textnormal{Error due to tails}.
\]
We can estimate each of these terms individually. \\ \ \\
\noindent\emph{Error between $\Gamma$ and $\Gamma_1$}:  It follows from the definition of $\Gamma_1$ that
\begin{align*}
\big| \Gamma_1(\omega) - \Gamma(\omega)\big| &= \frac{\kappa}{\pi}\bigg|\int_{-\omega_c}^{\omega_c}\frac{\Gamma(\nu)}{(\omega - \nu)^2+\kappa^2}d\nu - \int_{-\infty}^{\infty}\frac{\Gamma(\omega)}{(\omega - \nu)^2 + \kappa^2}d\nu \bigg| \nonumber\\
&\leq \frac{\gamma}{\pi} \int_{|\nu|\geq \omega_c} \bigg| \frac{\Gamma(\omega)}{(\omega - \nu)^2 + \kappa^2}\bigg| d\nu +  \frac{\kappa}{\pi}\int_{-\omega_c}^{\omega_c}\bigg| \frac{\Gamma(\omega) - \Gamma(\nu)}{(\omega - \nu)^2+\kappa^2}\bigg| d\nu.
\end{align*}
The first integral can analytically integrated to obtain:
\begin{align*}
\frac{\gamma}{\pi}\int_{|\nu| \geq \omega_c} \bigg|\frac{\Gamma(\omega)}{(\omega - \nu)^2 + \kappa^2} \bigg|d\nu = \Gamma(\omega)  \bigg(1 -
 \frac{1}{\pi}\tan^{-1}\frac{|\omega - \omega_c|}{\kappa} - \frac{1}{\pi}\tan^{-1}\frac{|\omega + \omega_c|}{\kappa}\bigg).
\end{align*}
To estimate the second integral, we make use of Taylor's theorem: $ \forall \nu \in [-\omega_c, \omega_c]: \ |\Gamma(\omega) - \Gamma(\nu)| \leq \Gamma_\textnormal{max}^d(\omega_c) |\omega - \nu|$ where $\Gamma_\textnormal{max}^d(\omega_c) = \sup_{\omega \in [-\omega_c, \omega_c]}|\Gamma'(\omega)|$. It then follows that $\forall \omega \in [-\omega_c, \omega_c]$ that 
\begin{align*}
\frac{\kappa}{\pi}\int_{-\omega_c}^{\omega_c} \bigg|\frac{\Gamma(\omega) - \Gamma(\nu)}{(\omega - \nu)^2 + \kappa^2} \bigg| d\nu \leq \frac{\kappa\Gamma_\text{max}^d(\omega_c)}{2\pi} \bigg(\log \frac{(\omega - \omega_c)^2 + \kappa^2}{\kappa^2} + \log \frac{(\omega + \omega_c)^2 + \kappa^2}{\kappa^2}\bigg).
\end{align*}
We thus obtain that
\[
\int_{-\omega_c}^{\omega_c} |\Gamma_1(\omega) - \Gamma(\omega) |d\omega\leq \omega_c\Gamma_\text{max}(\omega_c) \xi\bigg(\frac{\kappa}{\omega_c}\bigg) + \omega_c^2\Gamma_\text{max}^d \xi^d\bigg(\frac{\kappa}{\omega_c}\bigg)
\]
where
\begin{align*}
&\xi(x) = \frac{1}{\pi} \tan^{-1}\bigg(\frac{x}{2}\bigg) - \frac{x}{\pi}\log\bigg(1 + \frac{4}{x^2}\bigg), \nonumber \\
&\xi^d(x) = 4x \log\bigg(1 + \frac{4}{x^2}\bigg) - 4x + 4x \tan^{-1}\bigg(\frac{2}{x}\bigg).
\end{align*}
We note that $\xi, \xi_d = O(x \text{log}(1 / x))$ and therefore
\[
\int_{-\omega_c}^{\omega_c}\big|\Gamma_1(\omega) - \Gamma(\omega)\big|d\omega \leq O\bigg(\kappa \Gamma_\text{max}\log\frac{\kappa}{\omega_c}\bigg) + O\bigg(\omega_c \kappa \Gamma_\text{max}^d(\omega_c) \log\frac{\kappa}{\omega_c}\bigg)
\]

\noindent\emph{Error between $\Gamma_1$ and $\hat{\Gamma}$}: Next, to obtain $\hat{\Gamma}$ from $\Gamma_1$, we simply discretize the integral in the definition of $\Gamma_1$ --- the Taylor's theorem can again be used to bound the resulting discretization error. In particular,
\[
\int_{-\omega_c}^{\omega_c} \big|\Gamma_1(\omega) - \hat{\Gamma}(\omega)\big | d\omega \leq \frac{2\omega_c^3}{\pi (M - 1)}\bigg(\frac{\Gamma_\text{max}^d(\omega_c)}{\kappa} + \frac{3\sqrt{3}}{8}\frac{\Gamma_\text{max}(\omega_c)}{\kappa^2}\bigg) = O\bigg(\frac{\omega_c^3 \Gamma_\text{max}^d(\omega_c)}{\kappa M}\bigg) + O\bigg(\frac{\omega_c^3 \Gamma_\text{max}}{\kappa^2 M} \bigg).
\]

\noindent\emph{Error due to tails:} This error can be explicitly evaluated.
\begin{align*}
\int_{|\omega|\geq \omega_c} |\hat{\Gamma}(\omega)|d\omega &\leq \sum_{n = 1}^{M}\frac{\kappa \delta \Gamma_\text{max}}{\pi} \int_{|\omega|\geq\omega_c} \frac{1}{(\omega - \omega_n)^2 + \kappa^2}d\omega, \nonumber\\
&= \sum_{n=1}^M \frac{\delta \Gamma_\text{max}}{\pi}\bigg(\tan^{-1}\bigg(\frac{\kappa}{\omega_c - \omega_n}\bigg)+\tan^{-1}\bigg(\frac{\kappa}{\omega_c + \omega_n}\bigg)\bigg),\nonumber\\
&=\delta \Gamma_\text{max} + \frac{2\delta \Gamma_\text{max}}{\pi}\sum_{n=1}^{M - 1}\tan^{-1}\bigg(\frac{\kappa}{n \delta}\bigg).
\end{align*}
Noting that for $x > 0$, $\tan^{-1}(1/x)$ is convex upwards, it follows that for any $\theta > 0$
\[
\sum_{n=1}^{M-1}\tan^{-1}\bigg(\frac{1}{n\theta}\bigg) \leq \frac{1}{\theta} \int_0^{(M - 1)\theta}\tan^{-1}\bigg(\frac{1}{x}\bigg) dx = (M - 1)\tan^{-1}\bigg(\frac{1}{(M - 1)\theta}\bigg) + \frac{1}{2\theta}\log\big(1 + (M - 1)^2\theta^2\big).
\]
Consequently,
\[
\int_{|\omega|\geq\omega_c}|\hat{\Gamma}(\omega)|d\omega \leq O\bigg(\frac{\Gamma_\text{max}}{M}\bigg)  + O\bigg(\kappa \Gamma_\text{max}\log\frac{\kappa}{\omega_c}\bigg).
\]
This proves the lemma $\square$.
 \\ \ \\
\textbf{Theorem 2 (Pseudomode approximation, restated)}
\emph{Suppose $v \in \textnormal{C}^\infty_b(\mathbb{R})$ is a coupling function such that assumptions 1 and 2 are satisfied and let $\rho(t)$ be the reduced state of the local system after evolving an initial state $\ket{\sigma}\otimes \ket{\textnormal{vac}}$ using the Hamiltonian in Eq.~1. Then, there exists a pseudomode description of the environment (definition 3) with $M$ bosonic modes which provides an approximation $\hat{\rho}(t)$ to the reduced state of the local system such that $\norm{\rho(t) - \hat{\rho}(t)}_\textnormal{tr} \to 0$ as $M \to \infty$. Furthermore, if $|v'(\omega)| = O(\textnormal{poly}(\omega))$ and the cutoff error $\varepsilon(\omega_c, t) = O(\exp(O(t))\textnormal{poly}(\omega_c^{-1}))$, then there exists a pseudomode description of the non-Markovian system with $M$ bosonic modes such that the trace-norm error in approximating the reduced local system state at time $t$ scales as $O(\exp(O(t))\textnormal{poly}(M^{-1}))$.} \\

\noindent\emph{Proof}: For a coupling function $ v\in \text{C}^\infty_b(\mathbb{R})$, we denote by $\rho(t)$ the true reduced state of the local system at time $t$ i.e. the reduced state obtained when evolved as per the Hamiltonian in Eq.~1 of the main text. First we introduce a frequency cutoff $\omega_c$ --- we denote by $\rho_{\omega_c}(t)$ the reduced state of the local system obtained when evolved as per Hamiltonian in Eq.~1 of the main text but with coupling function $v \cdot \text{rect}_{\omega_c}$. Second, we approximate $v\cdot \text{rect}_{\omega_c}$ by a sum of lorentzians by using the construction of lemma \ref{lemma:lemma_appx} to obtain a pseudomode approximation to the non-Markovian system (lemma \ref{lemma:pseudo_mode_appx}). Let $\hat{\rho}(t)$ be the reduced state of the local system at time $t$ obtained on using the pseudomode approximation. We estimate the error between the true reduced state $\rho(t)$ and the reduced state using the pseudomode approximation $\hat{\rho}(t)$ via
\[
\norm{\rho(t) - \hat{\rho}(t)}_\text{tr}\leq \underbrace{\norm{\rho(t) - \rho_{\omega_c}(t)}_\text{tr}}_\text{Cutoff error} + \underbrace{\norm{\rho_{\omega_c}(t) - \hat{\rho}(t)}_\text{tr}}_\text{Lorentzian approximation error}.
\]
From theorem 1, it follows that the cutoff error decreases as $\omega_c\to \infty$. To estimate the Lorentzian approximation error, we use lemmas \ref{lemma:final_appx_error} and \ref{lemma:lemma_appx}, which allows us to relate this error to the frequency cutoff $\omega_c$, number of pseudomodes $M$ as well as the linewidths of the pseudomodes $\kappa$. The key idea behind showing that the sum total of these two errors goes to 0 as $M \to \infty$ is to choose $\kappa$ and $\omega_c$ as a function of $M$ such that (i) the Lorentzian approximation error decreases with $M$ and (ii) $\omega_c$ increases with $M$ (as a consequence of which the cutoff error decreases with $M$). Below, we show that this choice can always be made for any coupling function $v$.

We make the choice $\kappa = \Theta(M^{-1/4})$. Furthermore, consider the increasing functions $f_1, f_2, f_3, f_4, f_5:[0, \infty) \to [0, \infty)$ of the cutoff frequencies that appear in the error estimate --- $f_1(\omega_c) = \log \omega_c$, $f_2(\omega_c) = \omega_c\Gamma_\text{max}^d(\omega_c)$, $f_3 = \omega_c \log(\omega_c)\Gamma_\text{max}^d(\omega_c)$, $f_4(\omega_c) = \omega_c^3$ and $f_5(\omega_c) = \omega_c^3 \Gamma_\text{max}^d(\omega_c)$. We note that since by definition $\Gamma_\text{max}^d(\omega_c)$ is an increasing non-negative function, $f_1, f_2, f_3, f_4 = O(f_5)$ as $\omega_c\to \infty$. We choose $\omega_c = \Theta(f_5^{-1}(M^{1/5}))$ --- we note that since $f_5^{-1}$ is an increasing function, $f_1(\omega_c), f_2(\omega_c), f_3(\omega_c), f_4(\omega_c) = O(M^{1/8})$ as $M\to \infty$. Using these choices of $\kappa$ and $\omega_c$, the error in approximating $|v|^2$ with a sum of lorentzian, as calculated in Lemma \ref{lemma:lemma_appx}, is $O( M^{-1/8})$ and thus goes to 0 as $M \to \infty$. Furthermore, since $\omega_c$ grows with $M$, it follows from theorem 1 that the cutoff error goes to 0 as $M \to\infty$.

To show that, under the assumption of a polynomial growth in $|v'(\omega)|$ with $\omega$ and a polynomial fall off of the cutoff error with $\omega$, the total approximation error decreases polynomially with the number of pseudomodes, we simply have to note that in the above proof, $f_5^{-1}(M^{1/5})$ will be $\text{poly}(M)$ and consequently the construction of theorem 3 yields a cutoff frequency $\omega_c$ which grows polynomially with $M$. It is already shown in the proof of theorem 3 that the error in approximating the magnitude square of the coupling function with a sum of Lorentzian reduces polynomially with the number of pseudomodes, and with this choice of $\omega_c$ the cutoff error also reduces polynomially with the number of pseudomodes $\square$.

\subsection{Proof of theorem 3: Convergence of star-to-chain transformation}\label{sec:detailed_proofs:lanczos}
\noindent\textbf{Definition 4 (Chain description, restated)}
\emph{An environment described by a chain of $M$ bosonic modes with parameters $\{(\omega_i, g_i) : i \in \{0, 1, 2, 3 \dots M - 1\}\}$ has an associated Hilbert space $\mathcal{H}_\textnormal{aux}$ of $M$ bosonic modes with annihilation operators $a_0, a_1 \dots a_{M - 1}$,
For a local system with Hilbert space $\mathcal{H}_S = \mathbb{C}^d$, time-dependent Hamiltonian $H_S(t) \in \mathfrak{L}(\mathbb{C}^d)$ interacting with the environment through the operator $L \in \mathfrak{L}(\mathbb{C}^d)$ with the initial system-environment state being $\ket{\sigma}\otimes \ket{\textnormal{vac}}$, its reduced state at time $t$ is given by $\rho(t) = \textnormal{Tr}_\textnormal{aux}(\hat{U}(t, 0)\ket{\sigma}\bra{\sigma} \otimes (\ket{0}\bra{0})^{\otimes M}\hat{U}(0, t))$, where $\hat{U}(\tau_1, \tau_2)$ is the propagator corresponding to the Hamiltonian
\[
\hat{H}_M(t) = H_S(t) + g_0 (L^\dagger a_0 + a_0^\dagger L) + H_{E, M},
\]
with
\[
H_{E, M} = \sum_{i = 0}^{M - 1}\omega_i a_i^\dagger a_i + \sum_{i = 1}^{M-2}g_i(a_i a_{i + 1}^\dagger + a_{i+1}a_i^\dagger).
\]}

\noindent\textbf{Algorithm 1 (Lanczos iteration for generating star-to-chain mapping, Refs.~\cite{chin2010exact, woods2014mappings})} \\ \ \\
\emph{INPUT}: A coupling function $v \in \textrm{C}_b^\infty(\mathbb{R})$, a cutoff frequency $\omega_c \in (0, \infty)$ and the number of modes $M \in \mathbb{N}$. \\
\emph{OUTPUT}:
\begin{itemize}
\item Parameters $\{(\omega_i, g_i) : i \in \{0, 1, 2, 3 \dots M - 1\}\}$ of a chain description of the environment, and \item Polynomials $\{p_0(\omega), p_1(\omega) \dots p_{M - 1}(\omega)\}$, where $\forall i \in \{0, 1, 2 \dots M - 1\}$, $p_i(\omega)$ is a polynomial of degree $i$ such that the annihilation operator of the $i^\text{th}$ bosonic mode in the chain description, $a_i$, is given by
\[
a_i = \frac{\int_{-\omega_c}^{\omega_c} v(\omega)p_i(\omega)d\omega}{\sqrt{\int_{-\omega_c}^{\omega_c} |v(\omega)|^2 p_i^2(\omega)d\omega}}.
\]
\end{itemize}
\emph{ALGORITHM}:
\begin{itemize}
\item Initialize $p_0(\omega) = 1, \beta_0 = 0$.
\item For $i \in \{1, 2 \dots M - 1\}$, set
\[
\omega_{i - 1} = \frac{\int_{-\omega_c}^{\omega_c} \omega p_{i-1}^2(\omega) |v(\omega)|^2 d\omega}{\int_{-\omega_c}^{\omega_c} p_{i - 1}^2(\omega) |v(\omega)|^2 d\omega}, \ p_i(\omega) = (\omega - \omega_{i - 1})p_{i - 1}(\omega) - \beta_{i - 1} p_{i-2}(\omega) \text{ and } \beta_i = \frac{\int_{-\omega_c}^{\omega_c} p_{i-1}^2(\omega) |v(\omega)|^2d\omega}{\int_{-\omega_c}^{\omega_c}p_i^2(\omega) |v(\omega)|^2d\omega},
\]
and $g_{i-1} = \sqrt{\beta_{i}}$.
\end{itemize}
We make two remarks about the Lanczos iteration described in the above algorithm. First, the cutoff frequency $\omega_c$ is introduced into the algorithm so as to ensure that the integrals involved in computing the chain parameters are well defined. This is not necessary if the coupling functions decays fast enough with frequency so as to ensure that the integrals exist \cite{woods2014mappings}. Second, the chain transformation generated by the Lanczos iteration with $M \to \infty$ exactly reproduces the dynamics of the local system interacting with an environment restricted to $[-\omega_c, \omega_c]$. However, only using a finite number of modes results in a truncation error --- in the following lemma, we first analyze this truncation error and then combine it with the error incurred with the introduction of the frequency cutoff.
\begin{lemma}\label{lemma:lanczos_error_int_pic}Let $\rho_{M\to\infty}(t)$ and $\rho_M(t)$ be the reduced state of the local system obtained on evolving an initial state $\ket{\sigma}\otimes \ket{\textnormal{vac}}$, with $\ket{\sigma} \in \mathcal{H}_S$, with respect to $\hat{H}_{M\to \infty}(t)$ and $\hat{H}_{M}(t)$ (defined in definition 4) respectively, then
\begin{align*}
&\norm{\rho_{M\to \infty}(t) - {\rho_{M}(t)}}_\textnormal{tr}\leq \sqrt{2}g(t) \bigg(\sup_{s\in [0, t]} \Big|[\delta B_{M}(s) , \delta B_{M}^\dagger(s)]\Big|\bigg)^{1/2},
\end{align*}
where
\begin{align*}
g(t) = \int_0^t \bigg(\sqrt{1 + \norm{v}_\infty^2 \tau \norm{L} ^2 e^{2\pi\norm{v}_\infty^2 \norm{L} ^2 \tau}} + \norm{v}_\infty \sqrt{\tau} \norm{L}  e^{\pi\norm{v}^2_\infty \norm{L} ^2 \tau }\bigg)d\tau,
\end{align*}
and
\[
\delta B_{M}(s) = g_0 \big(e^{isH_{E, M\to \infty} } a_0 e^{-isH_{E, M\to \infty}} - e^{isH_{E, M}} a_0 e^{-isH_{E, M}}\big),
\]
with $g_0, a_0$ and $H_{E, M}$ being given in definition 4.
\end{lemma}
\emph{Proof}: The proof of this lemma is very similar to that of lemma \ref{lemma:error_state}, but repeating the analysis in the interaction picture with respect to the Hamiltonian of the environment ($H_{E, M\to \infty}$ or $H_{E, M}$ for the two models under consideration). The two Hamiltonians $\hat{H}_{M\to \infty}(t)$ and $\hat{H}_M(t)$, after moving into the interaction picture, map to $\bar{H}_{M\to \infty}(t)$ and $\bar{H}_{M}(t)$ respectively where
\begin{align*}
&\bar{H}_{M\to \infty}(t) = e^{it H_{E, M\to \infty}t} \hat{H}_{M\to \infty}e^{-it H_{E, M\to \infty}t} = H_S(t) + \big(B_{M\to\infty}(t) L^\dagger + B^\dagger_{M\to\infty}(t) L \big), \\
&\bar{H}_{M}(t) = e^{it H_{E, M}t} \hat{H}_{M}e^{-it H_{E, M}t} =H_S(t) + \big(B_{M}(t) L^\dagger + B^\dagger_{M}(t) L\big),
\end{align*}
where $B_{M\to \infty}(t) = g_0 e^{it H_{E, M\to\infty}}a_0 e^{-it H_{E, M\to\infty}}$ and $B_{M}(t) = g_0 e^{it H_{E, M}} a_0 e^{-itH_{E, M}}$. We denote by $\ket{\psi_{M\to\infty}(t)}$ and $\ket{\psi_{M}(t)}$ as the solution of the Schroedinger equation with Hamiltonians $\tilde{H}_{M\to\infty}(t)$ and $\tilde{H}_{M}(t)$ and with the initial state being $\ket{\sigma}\otimes\ket{\text{vac}}$. We note that the norm of the $N-$particle component of $\ket{\psi_{M\to\infty}(t)}$ will satisfy the bound in lemma 2 of the main text since it corresponds to the quantum state obtained from a Hamiltonian of the form Eq.~1 of the main text with coupling function 
$v\cdot\text{rect}_{\omega_c}$, and translating this state into the interaction picture does not change these norms. Consequently, we obtain
\begin{align*}
&\norm{\ket{\psi_{M}(t)} - \ket{\psi_{M\to \infty}(t)}} \\&\qquad\leq \int_0^t \norm{(\bar{H}_{M}(s) - \bar{H}_{M\to\infty}(s))\ket{\psi_{\omega_c}(s)}}ds, \nonumber\\
&\qquad\leq\norm{L} \int_0^t  \bigg(\norm{\delta B_{M}(s) \ket{\psi_{M\to\infty}(s)}} + \norm{\delta B_{M}^\dagger(s) \ket{\psi_{M\to\infty}(s)}}\bigg)ds.
\end{align*}
Noting that for an $N-$particle wave-packet $\ket{\psi^N}$ and for all $s \in [0, t]$
\begin{align*}
&\norm{\delta B_{M}(s) \ket{\psi^N}}^2 \leq \norm{\ket{\psi^N}}^2 N \sup_{s\in[0, t]}\Big |[\delta B_{M}(s), \delta B_{M}^\dagger(s)] \Big |,  \\
&\norm{\delta B_{M}^\dagger(s) \ket{\psi^N}}^2 \leq \norm{\ket{\psi^N}}^2 (N + 1) \sup_{s\in [0, t]}\Big |[\delta B_{M}(s), \delta B_{M}^\dagger(s)] \Big |.
\end{align*}
Using this together with the bound in lemma 2 of the main text, we obtain that
\begin{align*}
&\norm{\ket{\psi_{M\to\infty}(t)} - \ket{\psi_{M}(t)}} \leq g(t)\bigg(\sup_{s\in[0, t]}\Big |[\delta B_{M}(s), \delta B_{M}^\dagger(s)] \Big |\bigg)^{1/2}.
\end{align*}
Finally, the lemma follows using the contractivity of the partial trace together with Eq.~\ref{eq:tr_norm_l2} $\square$.

An explicit dependence of the error bound in lemma \ref{lemma:lanczos_error_int_pic} on the number of modes $N$ can be obtained by utilizing the connection between orthogonal polynomials and the gauss quadrature method for approximating integrals with summations. This is provided in the following lemma.
\begin{lemma}\label{lemma:gauss_quad_error} Let $\delta B_{M}(s)$ be as defined in lemma \ref{lemma:lanczos_error_int_pic}, then
\[
\Big|[\delta B_{M}(s), \delta B_{M}^\dagger(s)]\Big| \leq O\bigg(\frac{\norm{v}_\infty^2 (2\omega_c)^{N + 2}s^{N + 1}}{(N - 1)!}\bigg).
\]
\end{lemma}
\noindent\emph{Proof}: We start by noting that $H_{E, M\to\infty} = \int_{-\omega_c}^{\omega_c}\omega a_\omega^\dagger a_\omega d\omega$ and therefore 
\[
g_0\big(e^{is H_{E, M\to\infty}}a_0 e^{-is H_{E, M \to \infty}}\big) = \int_{-\omega_c}^{\omega_c}v(\omega) a_\omega e^{-i\omega s}d\omega.
\]
Evaluating $g_0\big(e^{isH_{E, M}} a_0 e^{-is H_{E, M}}\big)$ is more involved since we need to diagonalize $H_{E, M}$. It is convenient to introduce the polynomial $\pi_i$, which is $p_i$ (defined in the algorithm 1) after normalization
\[
\pi_i(\omega) = \frac{p_i(\omega) \big(\int_{-\omega_c}^{\omega_c}|v(\omega)|^2d\omega\big)^{1/2}}{\big(\int_{-\omega_c}^{\omega_c}p_i^2(\omega) |v(\omega)|^2d\omega\big)^{1/2}}
\]
Suppose $\Phi = \sum_{i = 0}^{M - 1} \phi_i a_i$ is an eigenmode of $H_{E, M}$ at eigenfrequency $\Omega$ i.e. $[\Phi, H_{E, M}] = \Omega \Phi$. Using the explicit expression for $H_{E, M}$, we obtain that
\[
(\omega_i - \Omega) \phi_i + \sqrt{\beta_i}\phi_{i-1} + \sqrt{\beta_{i + 1}}\phi_{i + 1} = 0,
\]
where $\beta_i$ is defined in algorithm 1. It then follows that for $i \in \{0, 1, 2\dots N - 1\}$, $\phi_i = \pi_i(\omega) \phi_0$ and $p_M(\Omega) = 0$. Thus the possible eigenenergies of $H_{E, M}$, denoted by $\Omega_0, \Omega_1 \dots \Omega_{M - 1}$, are roots of the $M^\text{th}$ orthogonal polynomial $p_M$. For $i \in \{0, 1, 2\dots M - 1\}$, the eigenmode $\Phi_i$ associated with $\Omega_i$ can be written as $\Phi_i = \sum_{j = 0}^{M - 1}\phi^i_j a_j$ where
\begin{align}\label{eq:eigvec}
\phi_i^j = \frac{\pi_i(\Omega_j)}{N_j} \ \text{where } N_j = \bigg[\sum_{i=0}^{N - 1}\pi_i^2(\Omega_j)\bigg]^{1/2}.
\end{align}
We note that from the orthogonality of the eigenmodes $\phi_i$, it follows that
\begin{align}
\sum_{i = 0}^{N - 1} \pi_i(\Omega_j) \pi_i(\Omega_{j'}) = N_{j}^2\delta_{j, j'} \text{ and }\sum_{j=0}^{N - 1}\pi_i(\Omega_j) \pi_{i'}(\Omega_j) = N_j^2 \delta_{i, i'}.
\end{align}
Finally, consider evaluating $e^{isH_{E, M}} a_0 e^{-is H_{E, M}}$. Noting that $a_0 = \sum_{i = 0}^{M-1} \phi_0^j \Phi_j$, we obtain
\[
e^{isH_{E, M}} a_0 e^{-is H_{E, M}}  = \sum_{j = 0}^{N - 1} \phi_0^j \Phi_j e^{-i\Omega_j s}.
\]
Finally, we evaluate the commutator $[\delta B_M(s), \delta B_M^\dagger(s)]$ --- with straightforward computation, it follows that
\begin{align}\label{eq:commutator_expansion}
[\delta B_M(s), \delta B_M^\dagger(s)] = 2\int_{-\infty}^\infty |v(\omega)|^2 d\omega - 2\sum_{i, j=0}^{N - 1}\phi^j_0 \phi^j_i{\int_{-\omega_c}^{\omega_c}|v(\omega)|^2 \pi_i(\omega)\cos((\omega - \Omega_j)s)d\omega}
\end{align}
Next we use the Gauss quadrature theorem\cite{mcclarren2018gauss} --- we note that the polynomials $p_0, p_1, p_2 \dots $ generated in algorithm 1 are the same polynomials that would be generated if a function of the form $P(\omega)|v(\omega)|^2$ for some $P$ has to be integrated in the interval $[-\omega_c, \omega_c]$ using Gaussian quadrature. The $M$ roots of $p_{M}$ correspond to the points within the interval $[-\omega_c, \omega_c]$ at which the integrand needs to be evaluated to compute the integral. It is well known that the gaussian quadrature with $M$ such points exactly evaluates the integral if $P$ is a polynomial of degree $\leq 2M - 1$. More specifically, for a given $M > 0,\ \exists w_0, w_1 \dots w_{M - 1} > 0$ with $\sum_{i=0}^{M - 1}w_i = 1$ such that for all polynomials $P$ of degree $\leq 2M - 1$,
\[
\int_{-\omega_c}^{\omega_c} P(\omega) |v(\omega)|^2 d\omega = \bigg(\int_{-\omega_c}^{\omega_c}|v(\omega)|^2d\omega\bigg)\sum_{i = 0}^{M - 1}w_i P(\Omega_i).
\] 
To use this result, we note that from Taylor's theorem it follows that the cosine in Eq.~\ref{eq:commutator_expansion} can be expanded into the sum of a degree $M$ polynomial and a remainder i.e. $\forall j \in \{0, 1, 2 \dots M-1\}, \forall \omega \in [-\omega_c, \omega_c]: \cos((\omega - \Omega_j)s) = C_M(\omega - \Omega_j) + R_M(\omega - \Omega_j)$ where $C_M$ is a degree $N$ polynomial with $C_M(0) = 1$ and $R_M$ is the remainder which satisfies the estimate $\norm{R_M}_\infty \leq (2\omega_c s)^{M + 1} / (M + 1)!$. With this decomposition of the cosine function, Eq.~\ref{eq:commutator_expansion} reduces to
\begin{align*}
&[\delta B_M(s), \delta B_M^\dagger(s)] \nonumber\\
&= 2\int_{-\infty}^\infty|v(\omega)|^2d\omega - 2\sum_{i, j = 0}^{M -  1} \phi_0^j \phi_i^j \int_{-\omega_c}^{\omega_c}|v(\omega)|^2\pi_i(\omega) C_M(\omega - \Omega_j) d\omega - 2\sum_{i, j = 0}^{M - 1}\phi_0^j \phi_i^j \int_{-\omega_c}^{\omega_c}|v(\omega)|^2 R_M(\omega - \Omega_j) \pi_i(\omega) d\omega.
\end{align*}
Noting that for $i \leq M - 1$, $\pi_i(\omega) C_M(\omega - \Omega_j)$ is a polynomial of degree $\leq 2M - 1$, we obtain by an application of the Gauss quadrature theorem and using Eq.~\ref{eq:eigvec} that
\begin{align*}
&\sum_{i, j = 0}^{M - 1}\phi_0^j \phi_i^j \int_{-\omega_c}^{\omega_c} |v(\omega)|^2\pi_i(\omega) C_M(\omega - \Omega_j) d\omega \nonumber\\
&\qquad=\bigg(\int_{-\omega_c}^{\omega_c}|v(\omega)|^2d\omega\bigg)\sum_{i, j, k = 0}^{M - 1}w_k  \frac{\pi_i(\Omega_j)\pi_i(\Omega_k)}{N_j^2} C_M(\Omega_k - \Omega_j)\nonumber\\
&\qquad=\bigg(\int_{-\omega_c}^{\omega_c}|v(\omega)|^2d\omega\bigg)\sum_{i, k = 0}^{M - 1}w_k \delta_{i, k} C_M(\Omega_k - \Omega_j) = \bigg(\int_{-\omega_c}^{\omega_c} |v(\omega)|^2d\omega\bigg) \sum_{k = 0}^{M - 1}w_k = \int_{-\omega_c}^{\omega_c}|v(\omega)|^2d\omega.
\end{align*}
Consequently, it follows that
\[
[\delta B_M(s), \delta B_M^\dagger(s)] = - 2\sum_{i, j = 0}^{M - 1}\phi_0^j \phi_i^j \int_{-\omega_c}^{\omega_c}|v(\omega)|^2 R_M(\omega - \Omega_j) \pi_i(\omega) d\omega.
\]
Noting $|\phi_i^j| \leq 1$ for $i, j \in \{0, 1, 2\dots M - 1\}$, we then obtain the following estimate
\begin{align*}
\Big| [\delta B_M(s), \delta B_M^\dagger(s)]\Big| &\leq 2 \sum_{i, j = 0}^{M - 1} \bigg| \int_{-\omega_c}^{\omega_c} |v(\omega)|^2 R_M(\omega - \Omega_j) \pi_i(\omega) d\omega \bigg| \nonumber\\
&\leq 2M\norm{v}_\infty \norm{R_M}_{\infty} \sum_{i = 0}^{M - 1} \bigg| \int_{-\omega_c}^{\omega_c} |v(\omega)| \pi_i(\omega) d\omega\bigg|\nonumber\\
&\leq 2M\norm{v}_\infty\frac{(2\omega_c s)^{M + 1}}{(M  + 1)!} \sum_{i = 0}^{M - 1} \bigg(2\omega_c \int_{-\omega_c}^{\omega_c}|v(\omega)|^2 \pi_i^2(\omega) d\omega\bigg)^{1/2}\nonumber\\
&\leq 4\omega_c \norm{v}_\infty^2 \frac{(2\omega_c s)^{M + 1}}{(M - 1)!} = O\bigg(\frac{\norm{v}_\infty^2 (2\omega_c)^{M + 2} s^{M + 1}}{(M - 1)!}\bigg),
\end{align*}
wherein in the last step we have used the fact that the polynomial $\pi_i(\omega)$ is normalized. This completes the proof $\square$.

\noindent\textbf{Theorem 3 (Star-to-chain transformation, restated)}\emph{
Suppose $v \in \textnormal{C}^\infty_b(\mathbb{R})$ is a coupling function such that assumptions 1 and 2 are satisfied and let $\rho(t)$ be the reduced state of the local system after evolving an initial state $\ket{\sigma}\otimes \ket{\textnormal{vac}}$ using the Hamiltonian in Eq.~1. Then, there exists a chain description of the environment with $M$ modes (definition 4) that provides an approximation $\hat{\rho}(t)$ to the reduced state of the local system such that $\norm{\rho(t) - \hat{\rho}(t)}_\textnormal{tr} \to 0$ as $M \to \infty$. Furthermore, if the cutoff error $\varepsilon(\omega_c, t) = O(\exp(O(t))\textnormal{poly}(\omega_c^{-1}))$, then the trace-norm error in approximating the reduced local system state at time $t$ scales as $O(\exp(O(t))\textnormal{poly}(M^{-1}))$.}

\noindent\emph{Proof}: The proof of this theorem parallels that of theorem 2. For a coupling function $ v\in \text{C}^\infty_b(\mathbb{R})$, we denote by $\rho(t)$ the true reduced state of the local system at time $t$ i.e. the reduced state obtained when evolved as per the Hamiltonian in Eq.~1 of the main text. First we introduce a frequency cutoff $\omega_c$ --- we denote by $\rho_{\omega_c}(t)$ the reduced state of the local system obtained when evolved as per Hamiltonian in Eq.~1 but with a frequency cutoff. Second, we use the Wilson's star-to-delta transformation with $M$ modes to obtain an approximation $\hat{\rho}(t)$ to $\rho_{\omega_c}(t)$. We estimate the error between the true reduced state $\rho(t)$ and the reduced state using the pseudomode approximation $\hat{\rho}(t)$ via
\[
\norm{\rho(t) - \hat{\rho}(t)}_\text{tr}\leq \underbrace{\norm{\rho(t) - \rho_{\omega_c}(t)}_\text{tr}}_\text{Cutoff error} + \underbrace{\norm{\rho_{\omega_c}(t) - \hat{\rho}(t)}_\text{tr}}_\text{Wilson star-to-delta error}.
\]
From theorem 1, it follows that the cutoff error decreases as $\omega_c\to \infty$. To estimate $\norm{\rho_{\omega_c}(t) - \hat{\rho}(t)}_\text{tr}$, we use lemmas \ref{lemma:lanczos_error_int_pic} and \ref{lemma:gauss_quad_error}, which allows us to relate this error to the frequency cutoff $\omega_c$ and the number of modes $M$. We note that choosing $\omega_c$ to increase sublinearly with $M$, both the cutoff error $\norm{\rho(t) - \rho_{\omega_c}(t)}_\text{tr}$ and $\norm{\rho_{\omega_c}(t) - \hat{\rho}(t)}_\text{tr}$ go to 0 as $M \to \infty$, thus providing a convergence guarantee for Wilson star-to-delta transformation. Furthermore, with this choice of $\omega_c$ and under the assumption that the cutoff error falls polynomially with $\omega_c$, it immediately follows that the approximation error decreases polynomially with the number of modes $M$ $\square$.

\edit{\section{Theorem 1 for an initially excited environment}
\label{sec:initial_env}
In this section, we consider the setting where the environment is initially in an excited state. Some minor modifications to the formulations of assumptions 1 and 2 allows us to extend theorem 1 to an initially excited environment state. We consider an initially mixed environment state, described as an ensemble $\rho_E(0) := \{p(\text{E}), \ket{\text{E}} \in \mathcal{H}_E\}$. We first assume that this ensemble is itself approximable within a finite frequency window, as is made precise below. \\ \ \\
\noindent\textbf{Assumption 3:} \emph{The initial environment state $\rho_E(0) = \{p(\ket{\textnormal{E}}, \ket{\textnormal{E}}\in \mathcal{H}_E\}$ is such that there is a function $\Delta(\omega_c)$ that vanishes as $\omega_c\to \infty$ and
\[
\underset{\ket{\textnormal{E}}}{\textnormal{esssup}} \norm{(\textnormal{id} - \Pi_{\omega_c}) \ket{\textnormal{E}}} \leq \Delta(\omega_c).
\]
where $\Pi_{\omega_c}$ is defined in definition \ref{def:projector}.
}

Furthermore, a physically reasonable environment state should modify assumption 2 to bound the derivative of Green's function over the pure states in the ensemble describing the mixed environment state.\\

\noindent\textbf{Assumption 2$'$} \emph{Given the initial environment state $\rho_E(0) = \{p(\ket{\textnormal{E}}), \ket{\textnormal{E}} \in \mathcal{H}_E\}$, $\forall N \in \mathbb{N}, t \geq 0, s \in [0, t]^{N - 1}$, the map $G_N^{\ket{\textnormal{E}}}(\{\cdot, s\}; t) : [0, t] \to \mathfrak{L}(\mathcal{H}_S)$ is absolutely continuous for almost all $\ket{\textnormal{E}}$ and $\exists \gamma(t) > 0$ such that
\[
\underset{\substack{\ket{\textnormal{E}} \sim \rho_E(0), \\ \{s, s_0\}\in[0, t]^N }}{\textnormal{esssup}} \norm{\partial_{s_0} G_N^\textnormal{E}(\{s_0, s\}; t)} \leq \norm{L}^N \gamma(t).
\]
}

\noindent\textbf{Theorem 1$'$} \emph{Suppose $v \in \textnormal{C}_b^\infty(\mathbb{R})$ is a coupling function and $\rho_E(0)$ is an initial environment state such that assumptions 1, 2$'$ and 3 are satisfied. Denoting by $\rho_S(t)$ the reduced density matrix of the local system at time $t$ when an initial state $\rho_S(0)\otimes\rho_E(0)$ is evolved under the Hamiltonian in Eq.~1 of the main text and by ${\rho}_{S, \omega_c}(t)$ the reduced density matrix of the local system at time $t$ when the same initial state is evolved under the Hamiltonian
\begin{align}\label{eq:hamiltonian_with_cutoff}
H_{\omega_c}(t) =& H_S(t) + \int_{-\infty}^{\infty} \omega a^\dagger_\omega a_\omega d\omega + \int_{-\omega_c}^{\omega_c} \bigg( v(\omega) a_\omega L^\dagger + v^*(\omega) a^\dagger_\omega L\bigg)d\omega,
\end{align}
then
\begin{align*}
\norm{\rho(t) - {\rho}_{\omega_c}(t)}_\textnormal{tr} \leq \varepsilon(\omega_c, t),
\end{align*}
where $\varepsilon(\omega_c, t)$ is the cutoff error given by
\begin{align}
\varepsilon(\omega_c, t) =\sqrt{2} \Delta(\omega_c) + \frac{f_1(t)}{\sqrt{\omega_c}} + \int_0^t f_2(\tau) \sqrt{V(\omega_c, \tau)}d\tau.
\end{align}
Here $V(\omega_c, t)$ is defined in assumption 1 and
\begin{align*}
&f_1(\tau)= \sqrt{2}\norm{v}_\infty\norm{L} ( 2+ \gamma(\tau) \tau) e^{\pi\norm{v}_\infty^2 \norm{L}^2\tau}, \\
&f_2(\tau) = \sqrt{2}\norm{L}^2 \tau (1 + \gamma(\tau) \tau) e^{\pi \norm{v}_\infty^2 \norm{L}^2 \tau },
\end{align*}
where $\gamma(t)$ is introduced in assumption 2$'$.} \\

\noindent\emph{Proof}: Suppose that $\rho_E(0) = \{p(\ket{\textrm{E}}), \ket{\textrm{E}} \in \mathcal{H}_E\}$, is the initial environment state, and suppose $\rho_S(0) = \sum_{i=1}^d p_i \ket{\sigma_i}\bra{\sigma_i}$ for some $\ket{\sigma_1}, \ket{\sigma_2}\dots \ket{\sigma_d} \in \mathcal{H}_S$. The initial joint state of the system and the environment can then be expressed as
\[
\rho(0) = \sum_{i = 1}^d \int p_i dp(\ket{\textrm{E}})  \ket{\sigma_i}\bra{\sigma_i} \otimes \ket{\textrm{E}}\bra{\textrm{E}}.
\]
We will first analyze the cutoff error obtained with an initial state $\ket{\sigma_i} \otimes \ket{\textrm{E}}$ for some $i \in \{1, 2 \dots d\}$ and $\ket{\textrm{E}} \sim \rho_E(0)$. We note that for this problem, lemmas \ref{lemma:decomp}, \label{lemma:error_direct} and \label{lemma:error_coup} are applicable if assumptions 1, 2$'$ and 3 hold, and we obtain that
\begin{align}
\label{eq:individual_ev_state_norm}
&\norm{U(t, 0) \ket{\sigma_i} \otimes \ket{\text{E}} - U_{\omega_c}(t, 0) \ket{\sigma_i}\otimes \ket{\text{E}}} 
\nonumber\\
&\qquad \qquad \leq \Delta(\omega_c) + \frac{\norm{v}_\infty \norm{L}}{\sqrt{\omega_c}} (2 + \gamma(t) t) e^{\pi \norm{v}_\infty^2 \norm{L}^2 t} + \int_0^t  \sqrt{V(\omega_c, \tau)}  \norm{L}^2(1 + \gamma(\tau) \tau) e^{\pi \norm{v}_\infty^2 \norm{L}^2\tau}d\tau
\end{align}
holds for all $i \in \{1, 2 \dots d\}$ and for almost all $\ket{\textrm{E}}$ drawn from the ensemble $\rho_E(0)$. Thus, we immediately obtain that
\begin{align}
\label{eq:full_env_state_norm}
&\norm{U(t, 0) \rho(0) U^\dagger(t, 0) - U_{\omega_c}(t, 0)\rho(0) U_{\omega_c}^\dagger(t, 0)}_\text{tr}  \nonumber\\
&\qquad \qquad \leq \sum_{i=1}^d \int p_i dp(\ket{\textrm{E}}) \norm{U(t, 0) \ket{\sigma_i}\bra{\sigma_i}\otimes \ket{\textrm{E}}\bra{\textrm{E}} U^\dagger(t, 0) - U_{\omega_c}(t, 0)\ket{\sigma_i}\bra{\sigma_i}\otimes \ket{\textrm{E}}\bra{\textrm{E}}  U_{\omega_c}(t, 0)}\leq  \nonumber \\ 
&\qquad \qquad \leq \sqrt{2}  \sum_{i=1}^d \int p_i dp(\ket{\textrm{E}})\norm{U(t, 0) \ket{\sigma_i}\otimes \ket{\textrm{E}}- U_{\omega_c}(t, 0)\ket{\sigma_i}\otimes \ket{\textrm{E}} }
\end{align}
From Eqs.~\ref{eq:individual_ev_state_norm} and \ref{eq:full_env_state_norm}, the lemma statement follows. $\square$.
}
\bibliography{references.bib}